\documentclass[11pt]{article}
\usepackage{graphicx}
\usepackage{multirow}
\usepackage[authoryear]{natbib}
\usepackage{amsmath, amssymb}
\usepackage[utf8]{inputenc}
\usepackage{amsthm}
\usepackage{subfigure}
\usepackage{color}
\usepackage[T3,OT1]{fontenc}
\usepackage[colorlinks,bookmarksopen,bookmarksnumbered,citecolor=blue,urlcolor=green,hypertexnames=false]{hyperref}
\DeclareSymbolFont{tipa}{T3}{cmr}{m}{n}
\DeclareMathAccent{\invbreve}{\mathalpha}{tipa}{16}
\usepackage{setspace}
\usepackage{geometry}
\newtheorem{theorem}{Theorem}
\begin{document}
\title{On the estimation of variance parameters in non-standard generalised linear mixed models: Application to penalised smoothing \footnote{This is a pre-print of an article published in \textit{Statistics and Computing}. The final authenticated version is available online at: \url{https://doi.org/10.1007/s11222-018-9818-2}.}}%
\author{Mar\'ia Xos\'e Rodr\'iguez - \'Alvarez$^{1,2}$, Maria Durban$^{3}$, Dae-Jin Lee$^{1}$, Paul H. C. Eilers$^{4}$\\\\      
		\small{$^{1}$ BCAM - Basque Center for Applied Mathematics}\\
		\small{Alameda de Mazarredo, 14. E-48009 Bilbao, Basque Country, Spain}\\
        \small{\texttt{mxrodriguez@bcamath.org}}\\
		\small{$^{2}$ IKERBASQUE, Basque Foundation for Science, Bilbao, Spain}\\
        \small{$^{3}$ Department of Statistics and Econometrics, Universidad Carlos III de Madrid, Legan\'es, Spain}\\
        \small{$^{4}$ Erasmus University Medical Centre, Rotterdam, the Netherlands}}
\maketitle
\date{}
\sloppy
\begin{abstract}
We present a novel method for the estimation of variance parameters in generalised linear mixed models. The method has its roots in \cite{Harville77}'s work, but it is able to deal with models that have a precision matrix for the random-effect vector that is linear in the inverse of the variance parameters (i.e., the precision parameters). We call the method SOP (Separation of Overlapping Precision matrices). SOP is based on applying the method of successive approximations to easy-to-compute estimate updates of the variance parameters. These estimate updates have an appealing form: they are the ratio of a (weighted) sum of squares to a quantity related to effective degrees of freedom. We provide the sufficient and necessary conditions for these estimates to be strictly positive. An important application field of SOP is penalised regression estimation of models where multiple quadratic penalties act on the same regression coefficients. We discuss in detail two of those models: penalised splines for locally adaptive smoothness and for hierarchical curve data. Several data examples in these settings are presented.
\end{abstract}
\section{Introduction}
The estimation of variance parameters is a statistical problem that has received extensive attention for more than 50 years. It originated with the ANOVA methodology proposed by Fisher in the 1920's, where estimates where obtained equating mean squared error to its expected value. However, the results yielded by this method were not optimal in some situations, for example, in the case of unbalanced data. Later on, \cite{Crump51} applied maximum likelihood (ML) under the assumption of normally distributed errors and random effects. But it was not until the 1970's when the estimation of variance parameters based on ML methods gained interest. The method of \emph{Restricted Maximum Likelihood} (REML) \citep{Patterson71} gave a solution to the problem of biased estimators of the variance parameters. However, one of the main obstacles to the use of this technique, at the time, was the fact that the calculation of ML/REML estimates requires the numerical solution of a non-linear problem. \cite{Patterson71} proposed an iterative solution using the Fisher Scoring algorithm, but it was \cite{Harville77} who proposed the first numerical algorithm to compute REML estimates of the variance parameters. His proposal is the inspiration of our work.

Along the years, several computational approaches have appeared with the aim of improving the computational burden of solving the score equations for the variance parameters: \cite{Smith1990} proposed the use of the EM algorithm, \cite{Graser87} suggested the use of the simplex algorithm to obtain the estimates directly from the likelihood, and \cite{Gilmour1995} developed a method based on the use of an average information matrix. 

In the context of Generalised Linear Mixed Models (GLMMs), estimation based on iterative re-weighted REML has been proposed independently by a number of authors \citep[e.g.,][]{Schall1991,Engel1994}, as an extension of the iterative re-weighted least squares algorithm for Generalised Linear Models \citep[GLM,][]{McCullagh89}. \cite{Breslow1993} proposed a general method based on Penalised Quasi-Likelihood (PQL) for the estimation of the fixed and random effects, and pseudo-likelihood for the variance parameters. As noted by \cite{Engel1996}, the estimation procedures discussed in all these papers are equivalent, although motivated from quite different starting points. 

The majority of the methods mentioned above impose a strong restriction on the vector of random effects: its variance-covariance matrix has to be linear in the variance parameters. The results we present in this paper relax that assumption to the case in which the linearity in the parameters is necessary on the precision matrix and not on the variance-covariance matrix. Our contribution is motivated by the need to estimate smoothing parameters in the context of penalised regression models with non-standard quadratic penalties.

Penalised spline regression \citep[P-splines,][]{Eilers1996} has become a popular method for estimating models in which the mean response (or linear predictor in the non-Gaussian case) is a smooth unknown function of one or more covariates. The method is based on the representation of the smooth component in terms of basis functions, and the estimation of the parameters by modifying the likelihood with a quadratic penalty on the coefficients. The size of the penalty is controlled by the so-called \emph{smoothing parameter}. The connection between penalised smoothing and linear mixed models was first established a long time ago \citep{Green87}, and it has become of common use in the last 15 years \citep{Currie2002, Currie2006, Lee2010, Wand2003}. The key point of the equivalence is that the smoothing parameter becomes the ratio between two variance parameters and, therefore, the methods mentioned above can be used to estimate directly the amount of smoothing needed in the model, instead of using methods based on the optimisation of some criteria such as Akaike Information Criteria (AIC) or Generalised cross-validation (GCV) \citep{Eilers1996, Wood2008}. Standard methods based on REML/ML can be applied when simple penalties are used, i.e., each regression coefficient is affected by a single penalty (by a single smoothing parameter). However, in some circumstances, the penalties present an overlapping structure, with the same coefficients being penalised simultaneously by several smoothing parameters. This includes important cases such as multidimensional penalised splines with anisotropic penalties or adaptive penalised splines. Estimation methods that can deal with this situation have been proposed in the smoothing literature \cite [e.g.,][]{Wood2011, Wood2016}, but they have the drawback of being very computationally demanding, especially when the number of smoothing parameters is large.

This work addresses this problem and presents a fast method for estimating the variance parameters/smoothing parameters in generalised linear mixed models/generalised additive models. The method can be used whenever the precision matrix of the random component (or the penalty matrix of the P-spline model) is a linear combination defined over the inverse of the variance parameters (smoothing parameters). We obtain simple expressions for the estimates of the variance parameters that are ratios between a sum of squares and a quantity related to the notion of effective degrees of freedom in the smoothing context \citep{Hastie1990}. We show the sufficient and necessary conditions that guarantee the positiveness of these estimates, and discuss several situations where these conditions can be easily verified. Particular cases of the method presented here have been introduced in \cite{MXRA2015}, which solved the problem in the case of anisotropic multidimensional smoothing, and in \cite{MXRA2015b}, where results for adaptive P-splines were first discussed. More recently, \cite{Wood2017} extended the abovementioned works to more general penalised spline models. The proposal discussed here presents two main advantages with respect to \cite{Wood2017}'s approach. First, the smoothing/variance parameter estimates described in \cite{Wood2017} rely on Moore-Penrose pseudoinverses of the penalty matrices, which, in our experience, may present numerical instabilities. Second, our proposal establishes an explicit connection between variance component estimates and effective degrees of freedom, which lacks in \cite{Wood2017}. Effective degrees of freedom are key components in smoothing models. They help in summarising a model, as partial effective degrees of freedom are measures of model components' complexity with strong intuitive appeal \cite[see, e.g.,][for an example in the agricultural field]{MX18}.

The rest of this paper is organised as follows: Section \ref{HfD} introduces the work by \cite{Harville77}, that constitutes the foundation of the work presented here. Section \ref{SOP} is the core of the paper: the new method, called SOP (Separation of Overlapping Precision matrices), is presented; and the connection between SOP and the notion of effective degrees of freedom is discussed. Section \ref{OP_WWW} describes several P-splines models whose estimation can be approached using SOP. We focus in this paper on adaptive P-splines and P-splines for hierarchical curve data. Illustrations with data examples are provided in Section \ref{examples}. A Discussion closes the paper. Some technical details have been added as Appendices. The estimating algorithm is detailed there.
\section{Estimation of variance parameters in generalised linear mixed models: \cite{Harville77}'s work and extensions}\label{HfD}
This Section is our little tribute to \cite{Harville77}'s paper, which was the inspiration for this work. \cite{Harville77}'s paper deals with ML/REML approaches to variance parameters estimation in linear mixed models (LMM) for Gaussian data. Nonetheless, estimation of GLMM can be done by repeated use of LMM methodology on a \textit{working} dependent variable \citep[see, e.g.,][where use is made of the results by \citealp{Harville77}]{Schall1991,Engel1994}. This is the approach we follow in this paper. 

Let $\boldsymbol{y} = \left(y_1,\ldots,y_n\right)^{\top}$ be a vector of $n$ observations. A GLMM can be written as 
\begin{equation}
g\left(\boldsymbol{\mu}\right) = \boldsymbol{X}\boldsymbol{\beta} + \boldsymbol{Z}\boldsymbol{\alpha},\;\;\mbox{with}\;\;\boldsymbol{\alpha}\sim N\left(\boldsymbol{0}, \boldsymbol{G}\right), 
\label{mm_equation_orig}
\end{equation}
where $\mu_i = E\left(y_i\mid \boldsymbol{\alpha}\right)$ and $g\left(\cdot\right)$ is the link function. The model assumes that, conditional on the random effects, $y_i$ is independently distributed with mean $\mu_i$ and variance $\mathbb{V}\mbox{ar}(y_i|\boldsymbol{\alpha}) = \phi\nu\left(\mu_i\right)$. Here, $\nu\left(\cdot\right)$ is a specified variance function, and $\phi$ is the dispersion parameter that may be known or unknown. In model (\ref{mm_equation_orig}), $\boldsymbol{X}$ and $\boldsymbol{Z}$ represent column-partitioned matrices, associated respectively with the fixed and random effects. We assume that $\boldsymbol{X}$ has full rank, $\boldsymbol{Z} = \left[\boldsymbol{Z}_{1},\ldots,\boldsymbol{Z}_{c}\right]$, and $\boldsymbol{\alpha} = \left(\boldsymbol{\alpha}_{1}^{\top},\ldots,\boldsymbol{\alpha}_{c}^{\top}\right)^{\top}$. Each $\boldsymbol{Z}_{k}$ corresponds to the design matrix of the $k$-th random component $\boldsymbol{\alpha}_{k}$, with $\boldsymbol{\alpha}_{k}$ being a $(q_{k} \times 1)$ vector ($k = 1, \ldots, c$). We assume further that $\boldsymbol{\alpha}_k\sim N\left(\boldsymbol{0}, \boldsymbol{G}_k\right)$ and that
\begin{equation*}
\boldsymbol{G} =  \bigoplus_{k = 1}^{c}\boldsymbol{G}_{k} = \bigoplus_{k = 1}^{c}\sigma_{k}^2\boldsymbol{I}_{q_k} = \mbox{diag}\left(\sigma_{1}^2\boldsymbol{I}_{q_1},\ldots,\sigma_{c}^2\boldsymbol{I}_{q_c}\right), 
\label{G_mm_equation}
\end{equation*}
where $\boldsymbol{I}_{m}$ is an identity matrix of order $m \times m$, and $\bigoplus$ denotes the direct sum of matrices. Note that the variance-covariance matrix $\boldsymbol{G}$ is linear in the variance parameters $\sigma_{k}^2$.

As noted before, estimation of model (\ref{mm_equation_orig}) can be approached by iterative fitting of a LMM  that involves a \textit{working} dependent variable $\boldsymbol{z}$ and a weight matrix $\boldsymbol{W}$ (updated at each iteration). The specific form of $\boldsymbol{z}$ and $\boldsymbol{W}$ is given in Appendix \ref{algorithm}. If $\phi$ and $\sigma_k^2$ ($k=1,\ldots,c$) are known, at each iteration, the updates for $\boldsymbol{\beta}$ and $\boldsymbol{\alpha}$ follow from the so-called Henderson equations \citep{Henderson1963} 
\begin{equation}
\underbrace{
\begin{bmatrix}
\boldsymbol{X}^{\top}{\boldsymbol{R}}^{-1}\boldsymbol{X} & \boldsymbol{X}^{\top}{\boldsymbol{R}}^{-1}\boldsymbol{Z} \\
\boldsymbol{Z}^{\top}{\boldsymbol{R}}^{-1}\boldsymbol{X} & \boldsymbol{Z}^{\top}{\boldsymbol{R}}^{-1}\boldsymbol{Z} + {\boldsymbol{G}}^{-1}
\end{bmatrix}}_{\boldsymbol{C}}
\begin{bmatrix} 
\boldsymbol{\widehat{\beta}}\\
\boldsymbol{\widehat{\alpha}}
\end{bmatrix}
=
\begin{bmatrix}
\boldsymbol{X}^{\top}{\boldsymbol{R}}^{-1}\boldsymbol{z}\\
\boldsymbol{Z}^{\top}{\boldsymbol{R}}^{-1}\boldsymbol{z}
\end{bmatrix},
\label{MX:linearsystem_1}
\end{equation}
which give rise to closed-form expressions
\begin{align}
\widehat{\boldsymbol{\beta}}  & = \left(\boldsymbol{X}^{\top}\boldsymbol{V}^{-1}\boldsymbol{X}\right)^{-1}\boldsymbol{X}^{\top}\boldsymbol{V}^{-1}\boldsymbol{z},\nonumber\\
\widehat{\boldsymbol{\alpha}}_k & = \boldsymbol{G}_k\boldsymbol{Z}_k^{\top}\boldsymbol{P}\boldsymbol{z}\;\;\;\;\;\;(k = 1, \ldots, c), 
\label{mm_randomest}
\end{align}
where $\boldsymbol{P} = \boldsymbol{V}^{-1} - \boldsymbol{V^{-1}}\boldsymbol{X}\left(\boldsymbol{X^{\top}V^{-1}X}\right)^{-1}\boldsymbol{X}^{\top}\boldsymbol{V^{-1}}$ with $\boldsymbol{V} = \boldsymbol{R} + \boldsymbol{Z}\boldsymbol{G}\boldsymbol{Z}^{\top}$ and $\boldsymbol{R} = \phi\boldsymbol{W}^{-1}$. The Henderson equations are of little use if the variances parameters $\phi$ and $\sigma_k^2$ ($k=1,\ldots,c$) are unknown. In his 1977 paper, Harville shows how to estimate them by REML by an elegant iterative algorithm. Let's first define
\[
\boldsymbol{T} = \left(\boldsymbol{I} + \boldsymbol{Z}^{\top}\boldsymbol{S}\boldsymbol{Z}\boldsymbol{G}\right)^{-1},
\]
where $\boldsymbol{S} = \boldsymbol{R}^{-1} - \boldsymbol{R}^{-1}\boldsymbol{X}\left(\boldsymbol{X}^{\top}\boldsymbol{R}^{-1}\boldsymbol{X}\right)^{-1}\boldsymbol{X}^{\top}\boldsymbol{R}^{-1}$. We note that $\boldsymbol{T}$ can be partitioned as follows
\[
\boldsymbol{T} = 
\begin{bmatrix}
\boldsymbol{T}_{11} & \boldsymbol{T}_{12} &  \cdots & \boldsymbol{T}_{1c}\\
\boldsymbol{T}_{21} & \boldsymbol{T}_{22} &  \cdots & \boldsymbol{T}_{2c}\\
\vdots & \vdots &  \ddots & \vdots\\
\boldsymbol{T}_{c1} & \boldsymbol{T}_{c2} &  \cdots & \boldsymbol{T}_{cc}\\
\end{bmatrix},
\]
where $\boldsymbol{T}_{ij}$ are matrices of order $q_i \times q_j$. In \cite{Harville77}, the updated estimate of $\sigma^2_k$ ($k = 1,\ldots,c$) is
\begin{equation}
\widehat{\sigma}_k^{2} = \frac{\widehat{\boldsymbol{\alpha}}_k^{\top[t]}\widehat{\boldsymbol{\alpha}}_k^{[t]}}{\mbox{ED}_k^{[t]}},
\label{var_per_component_harville}
\end{equation}
where
\begin{equation}
\mbox{ED}_k^{[t]} = q_k - \mbox{trace}\left(\boldsymbol{T}_{kk}^{[t]}\right),
\label{ed_per_component_harville}
\end{equation}
and  the superscript $[t]$ denotes quantities evaluated at current estimates of the variance parameters. From the estimates of $\boldsymbol{\beta}$ and $\boldsymbol{\alpha}$ follow an estimate for $\boldsymbol{z}$: $\widehat{\boldsymbol{z}} = \boldsymbol{X}\widehat{\boldsymbol{\beta}} + \boldsymbol{Z}\widehat{\boldsymbol{\alpha}}$. The residuals are $\boldsymbol{z} - \hat{\boldsymbol{z}}$. Harville uses 
\begin{equation}\label{e_sighar}
\hat{\phi} = \frac{\boldsymbol{z}^{\top}\boldsymbol{W}\left(\boldsymbol{z} - \widehat{\boldsymbol{z}}^{[t]}\right)}{n - \mbox{rank}(\boldsymbol{X})}, 
\end{equation}
to estimate the dispersion parameter (not always needed in GLMM). An alternative expression is \cite[see, e.g.,][]{Engel1990,MXRA2015}
\begin{equation}\label{e_sigwe}
\hat{\phi} = \frac{\left(\boldsymbol{z} - \widehat{\boldsymbol{z}}^{[t]}\right)^{\top}\boldsymbol{W}\left(\boldsymbol{z} - \widehat{\boldsymbol{z}}^{[t]}\right)}{n - \mbox{rank}(\boldsymbol{X}) - \sum_{k=1}^c\mbox{ED}_k^{[t]}}. 
\end{equation}
Here $\mbox{rank}(\boldsymbol{X}) + \sum_{k=1}^c\mbox{ED}_k^{[t]}$ can be interpreted as the effective model dimension. At convergence, eqns. (\ref{e_sighar}) and (\ref{e_sigwe}) give identical numerical values.
\subsection{Effective degrees of freedom in Harville's method}
As noted by \cite{Harville77}, the iterates derived from eqn. (\ref{var_per_component_harville}) have an intuitively appealing form. On each iteration, $\sigma_k^2$ is estimated by the ratio between the sum of squares of the estimates for $\boldsymbol{\alpha}_{k}$ and a number between zero and $q_k$. We now show that the denominator in eqn. (\ref{var_per_component_harville}) can in fact be interpreted as effective degrees of freedom in smoothing sensu, i.e., as the trace of a ``hat'' matrix \citep{Hastie1990}. 

First note that expression (\ref{mm_randomest}) reveals that $\boldsymbol{Z}_k\widehat{\boldsymbol{\alpha}}_k = \boldsymbol{Z}_k\boldsymbol{G}_k\boldsymbol{Z}_k^{\top}\boldsymbol{P}\boldsymbol{z}$. Thus, the ``hat'' matrix corresponding to the $k$-th random component $\boldsymbol{\alpha}_k$ is
\[
\boldsymbol{H}_k = \boldsymbol{Z}_k\boldsymbol{G}_k\boldsymbol{Z}_k^{\top}\boldsymbol{P},
\]
i.e., $\boldsymbol{H}_k\boldsymbol{z} = \boldsymbol{Z}_k\widehat{\boldsymbol{\alpha}}_k$. We now show that $\mbox{trace}\left({\boldsymbol{H}_k}\right) = \mbox{ED}_k$. It is easy to verify that 
\begin{equation}
\boldsymbol{T} = \left(\boldsymbol{I} + \boldsymbol{Z}^{t}\boldsymbol{S}\boldsymbol{Z}\boldsymbol{G}\right)^{-1} = \boldsymbol{G}^{-1}\left(\boldsymbol{G}^{-1} + \boldsymbol{Z}^{t}\boldsymbol{S}\boldsymbol{Z}\right)^{-1},
\label{T_C_rel}
\end{equation}
where $\left(\boldsymbol{G}^{-1} + \boldsymbol{Z}\boldsymbol{S}\boldsymbol{Z}^{t}\right)^{-1}$ is that partition of the inverse of $\boldsymbol{C}$ in (\ref{MX:linearsystem_1}) corresponding to the random vector $\boldsymbol{\alpha}$ \citep{Harville77, Johnson1995}. Exploiting the block structure of $\boldsymbol{Z}$ and $\boldsymbol{G}$, and making use of result (\ref{T_C_rel}) and (A4) in \cite{Johnson1995}, we have that
\begin{equation}
\boldsymbol{Z}_k^{\top}\boldsymbol{P}\boldsymbol{Z}_k\boldsymbol{G}_k  = \boldsymbol{I}_{q_k} - \boldsymbol{G}_k^{-1}\boldsymbol{C}^{*}_{kk} = \boldsymbol{I}_{q_k} - \boldsymbol{T}_{kk},
\label{j_t_result}
\end{equation}
where, to ease the notation, $\boldsymbol{C}^{*}$ denotes the inverse of $\boldsymbol{C}$, and $\boldsymbol{C}^{*}_{kk}$ denotes that partition of $\boldsymbol{C}^{*}$ corresponding to the $k$-th random component $\boldsymbol{\alpha}_k$. Thus,
\begin{align*} 
\mbox{trace}\left({\boldsymbol{H}_k}\right) & = \mbox{trace}\left(\boldsymbol{Z}_k\boldsymbol{G}_k\boldsymbol{Z}_k^{\top}\boldsymbol{P}\right) = \mbox{trace}\left(\boldsymbol{Z}_k^{\top}\boldsymbol{P}\boldsymbol{Z}_k\boldsymbol{G}_k\right)\\ & = \mbox{trace}\left(\boldsymbol{I}_{q_k} - \boldsymbol{T}_{kk}\right) = q_k - \mbox{trace}\left(\boldsymbol{T}_{kk}\right) \\ & = \mbox{ED}_k.
\label{ed_sap_2}
\end{align*}
\section{Separation of overlapping precision matrices: the SOP method}\label{SOP}
In previous Section we have discussed an estimating method for generalised linear mixed models where the variance-covariance matrix of the random component is linear in the variance parameters. However, more complex structures of the variance-covariance matrix appear in practice. The present research was motivated by our work on penalised spline regression. In spite of that, the method to be discussed in this Section is not confined to this area: it can be seen as a general estimating method for generalised linear mixed models with a precision matrix of a specific structure. As in Section \ref{HfD}, we consider the generalised linear mixed model
\begin{equation}
g\left(\boldsymbol{\mu}\right) = \boldsymbol{X}\boldsymbol{\beta} + \boldsymbol{Z}\boldsymbol{\alpha} = \boldsymbol{X}\boldsymbol{\beta} + \sum_{k=1}^{c}\boldsymbol{Z}_k\boldsymbol{\alpha}_k,
\label{mm_equation}
\end{equation}
with $\boldsymbol{\alpha}_k\sim N\left(\boldsymbol{0}, \boldsymbol{G}_k\right)$, $\boldsymbol{\alpha}\sim N\left(\boldsymbol{0}, \boldsymbol{G}\right)$, and $\boldsymbol{G} = \bigoplus_{k = 1}^{c}\boldsymbol{G}_{k}$. The main difference with respect to Section \ref{HfD} is that we do not assume that 
$\boldsymbol{G}_{k} = \sigma_k^2\boldsymbol{I}_{q_k}$, but we consider precision matrices of the form
\begin{equation}
\boldsymbol{G}_{k}^{-1} = \sum_{l = 1}^{p_k}\sigma_{k_l}^{-2}\boldsymbol{\Lambda}_{k_l},
\label{G_mm_equation_sop}
\end{equation}
where $\sigma_{k_l}^2$ ($l = 1,\ldots,p_k$ and $k = 1, \ldots, c$) are the variance parameters, and $\boldsymbol{\Lambda}_{k_l}$ are known symmetric positive semi-definite matrices of dimension $q_k \times q_k$. Note that we do not require $\boldsymbol{\Lambda}_{k_l}$ to be positive definite. The only requirement we need is that $\boldsymbol{G}_{k}^{-1}$ ($k = 1, \ldots, c$) are positive definite, and so are $\boldsymbol{G}^{-1}$ and its inverse, the variance-covariance matrix $\boldsymbol{G}$.

Expression (\ref{G_mm_equation_sop}) deserves some detailed discussion. Firstly, it is worth noting that we do not work with variance-covariance matrices, but with their inverses, the precision matrices. As said, the developments in this work have their origin on penalised spline methods. In Section~\ref{OP_WWW} the need to work with precision matrices will become clear, or, in the terminology of penalised splines, with penalty matrices. Secondly, what constitutes the main contribution of this paper is that we assume that each random component $\boldsymbol{\alpha}_k$ ($k=1,\ldots,c$) in model $(\ref{mm_equation})$ may be ``affected'' (shrunk) by several variance parameters. A particular case would be when $p_k = 1$ $\forall k$, in which case we are in the situation discussed in Section \ref{HfD}.

For the sake of simplicity, in some cases we will rewrite the precision matrix $\boldsymbol{G}^{-1}$ as follows
\begin{equation}
\boldsymbol{G}^{-1} = \sum_{l=1}^{p}\sigma_l^{-2}\widetilde{\boldsymbol{\Lambda}}_l,
\label{G_inv_compact}
\end{equation}
where $p = \sum_{k=1}^{c}p_k$. By a slight abuse of notation, let $\boldsymbol{\Lambda}_l$ denote the matrices involved in expression (\ref{G_mm_equation_sop}). The matrix $\widetilde{\boldsymbol{\Lambda}}_l$ is $\boldsymbol{\Lambda}_l$ padded out with zeroes. Some specific examples will be presented in Section \ref{OP_WWW} below. Expression (\ref{G_inv_compact}) makes it clear that the present work deals with the situation of generalised linear mixed models with a precision matrix for the random component that is linear in the precision parameters $\sigma_l^{-2}$. The next Section presents the proposed estimation method, that we call SOP.
\subsection{The SOP method}
Regardless of the structure of $\boldsymbol{G}$, estimation of $\boldsymbol{\beta}$ and $\boldsymbol{\alpha}$, and, when necessary, the dispersion parameter $\phi$, does not pose a problem, and can be done as discussed in Section \ref{HfD} above (see also Appendix \ref{algorithm} for a detailed description of the estimating algorithm). Recall that estimates for $\boldsymbol{\alpha}_k$ are obtained as $\widehat{\boldsymbol{\alpha}}_k = \boldsymbol{G}_k\boldsymbol{Z}_k^{\top}\boldsymbol{P}\boldsymbol{z}$. The hat matrix associated with the $k$-th random component $\boldsymbol{\alpha}_k$ is, once again, $\boldsymbol{H}_k = \boldsymbol{Z}_k\boldsymbol{G}_k\boldsymbol{Z}_k^{\top}\boldsymbol{P}$, and the effective degrees of freedom of this component is $\mbox{ED}_k = \mbox{trace}\left(\boldsymbol{H}_k\right)$. The variance parameters $\sigma_{k_l}^2$ ($l = 1,\ldots,p_k$ and $k = 1, \ldots, c$), however, cannot be estimated by means of Harville's approach. This is a consequence of $\boldsymbol{G}$ not being linear in the variance parameters.

The key of our approach is to work with $\boldsymbol{G}^{-1}$ instead of with $\boldsymbol{G}$. Given that $\boldsymbol{G}^{-1}$ is linear in the precision parameters $\sigma^{-2}_{k_l}$, the first-order partial derivatives of the (approximate) REML log-likelihood function can be explicitly obtained, as well as the REML-based estimates of the variance parameters.  We state the result in the following Theorem, whose proof is given in Appendix \ref{AppA}. 
\begin{theorem}
Let $\boldsymbol{G} = \bigoplus_{k = 1}^{c}\boldsymbol{G}_{k}$ be a symmetric positive definite matrix, with $\boldsymbol{G}^{-1}_k = \sum_{l=1}^{p_k}\sigma_{k_l}^{-2}\boldsymbol{\Lambda}_{k_l}$ symmetric positive definite and $\boldsymbol{\Lambda}_{k_l}$ known symmetric positive semi-definite. Then, the REML-based estimate updates of the variance parameters $\sigma_{k_l}^2$ ($l = 1,\ldots,p_k$ and $k = 1, \ldots, c$) are given by
\begin{equation}
\widehat{\sigma}^{2}_{k_l} = \frac{\widehat{\boldsymbol{\alpha}}_k^{{[t]}\top}\boldsymbol{\Lambda}_{k_l}\widehat{\boldsymbol{\alpha}}_k^{[t]}}{\mbox{ED}_{k_l}^{[t]}},
\label{var_per_component}
\end{equation}
where
\begin{equation}
\mbox{ED}_{k_l}^{[t]} = trace\left(\boldsymbol{Z}_k^{\top}\boldsymbol{P}^{[t]}\boldsymbol{Z}_k\boldsymbol{G}_k^{[t]}\frac{\boldsymbol{\Lambda}_{k_l}}{\widehat{\sigma}_{k_l}^{2[t]}}\boldsymbol{G}_k^{[t]}\right),
\label{ed_per_component}
\end{equation} 
with $\widehat{\boldsymbol{\alpha}}^{[t]}$, $\boldsymbol{P}^{[t]}$ and $\boldsymbol{G}^{[t]}$ evaluated at the current estimates $\widehat{\sigma}_{k_l}^{2[t]}$ ($l = 1,\ldots,p_k$ and $k = 1, \ldots, c$), and, when necessary, $\widehat{\phi}^{[t]}$.
\label{the1} 
\end{theorem}
Note that when $\boldsymbol{G}_k = \sigma_k^2\boldsymbol{I}_{q_k}$, expressions (\ref{var_per_component}) and (\ref{ed_per_component}) reduce to those of Harville (expressions (\ref{var_per_component_harville}) and (\ref{ed_per_component_harville}) respectively). 

An important and desirable property of the updates given in eqn. (\ref{var_per_component}) is that they are always nonnegative, provided that the previous estimates of the variance parameters are nonnegative. In addition, under rather weak conditions, these updates are strictly positive (although it
is possible to obtain values very close to zero).
\begin{theorem}
If $\widehat{\sigma}_{k_l}^{2[t]} > 0$, then the REML-based estimate updates of the variance parameters given in eqn. (\ref{var_per_component}) are larger or equal than zero, with strict inequality holding if: (i) $\mbox{rank}\left(\boldsymbol{X}, \boldsymbol{Z}_k\boldsymbol{G}_k^{[t]}\boldsymbol{\Lambda}_{k_l}\right) > \mbox{rank}\left(\boldsymbol{X}\right)$; and (ii) $\boldsymbol{z}$ (the \it{working} response vector) is not in the space spanned by the columns of $\boldsymbol{X}$.
\label{the2} 
\end{theorem}
The proof is provided in Appendix \ref{AppB}. We note that condition (i) is needed for both the numerator and denominator of eqn. (\ref{var_per_component}) to be strictly positive, while (ii) is only needed for the numerator. From an applied point of view, it would undoubtedly important to be able to check if the conditions are fulfilled before fitting the model. This may not be an easy task, since they depend on $\boldsymbol{G}_k^{[t]}$, and thus may vary from iteration to iteration. There are, however, common situations where condition (i) could be checked in advance:
\begin{itemize}
\item If $\boldsymbol{\Lambda}_{k_l}$ is of full rank, then condition (i) simplifies to $\mbox{rank}\left(\boldsymbol{X}, \boldsymbol{Z}_k\right) > \mbox{rank}\left(\boldsymbol{X}\right)$. We note that this condition is the same as that discussed by \cite{Harville77} in Lemma 1.
\item If $\boldsymbol{G}_k^{[t]}$ and $\boldsymbol{\Lambda}_{k_l}$ commute (i.e., $\boldsymbol{G}_k^{[t]}\boldsymbol{\Lambda}_{k_l}$ = $\boldsymbol{\Lambda}_{k_l}\boldsymbol{G}_k^{[t]}$), then condition (i) simplifies to $\mbox{rank}\left(\boldsymbol{X}, \boldsymbol{Z}_k\boldsymbol{\Lambda}_{k_l}\right) > \mbox{rank}\left(\boldsymbol{X}\right)$. Examples when $\boldsymbol{G}_k^{[t]}$ and $\boldsymbol{\Lambda}_{k_l}$ commute include, for instance, when both are diagonal.
\end{itemize}
We discuss these situations in more detail in Section \ref{OP_WWW}, where some examples of application of the SOP method are presented. 
\subsection{Effective degrees of freedom in the SOP method}
In line with the Harville method discussed in Section \ref{HfD}, the denominator of eqn. (\ref{var_per_component}) has been denoted as $\mbox{ED}_{\{\cdot\}}$, from effective degrees of freedom. Result (\ref{ed_per_component}) makes it easy to show that the sum of the $\mbox{ED}_{k_l}$ over the $p_k$ variance parameters involved in $\boldsymbol{G}^{-1}_k$ (see  eqn. (\ref{G_mm_equation_sop})) corresponds to $\mbox{ED}_k$ (the effective degrees of freedom of $\boldsymbol{\alpha}_k$)

\begin{align*}
\sum_{l=1}^{p_k}\mbox{ED}_{k_l} = \sum_{l=1}^{p_k}trace\left(\boldsymbol{Z}_k^{\top}\boldsymbol{P}\boldsymbol{Z}_k\boldsymbol{G}_k\frac{\boldsymbol{\Lambda}_{k_l}}{\sigma_{k_l}^{2}}\boldsymbol{G}_k\right) = trace\left(\boldsymbol{Z}_k^{\top}\boldsymbol{P}\boldsymbol{Z}_k\boldsymbol{G}_k\right) = trace\left(\boldsymbol{H}_k\right) = \mbox{ED}_k.
\end{align*}

As a consequence, at convergence, the estimated effective degrees of freedom associated with each random component in model (\ref{mm_equation}) is obtained as a by-product of the SOP method.

To finish this part, we would like to point out an interesting link between the upper bound for $\mbox{ED}_{k_l}$ (denoted with $\mbox{ED}^{ub}_{k_l}$) and condition (i) in Theorem \ref{the2}. It can be shown that
\[
\mbox{ED}_{k_l} \leq \mbox{rank}\left(\boldsymbol{X}, \boldsymbol{Z}_k\boldsymbol{G}_k\boldsymbol{\Lambda}_{k_l}\right) - \mbox{rank}\left(\boldsymbol{X}\right) = \mbox{rank}\left((I_n - \boldsymbol{P}_{\boldsymbol{X}})\boldsymbol{Z}_k\boldsymbol{G}_k\boldsymbol{\Lambda}_{k_l}\right) = \mbox{ED}^{ub}_{k_l},
\]
where $\boldsymbol{P}_{\boldsymbol{X}} = \boldsymbol{X}\left(\boldsymbol{X}^{\top}\boldsymbol{X}\right)^{-1}\boldsymbol{X}^{\top}$. Thus, if condition (i) in Theorem \ref{the2} is not verified, $\mbox{ED}_{k_l}$ would be exactly zero. We omit the proof of the previous result. It can be obtained in a similar fashion as in the paper by \cite{Cui2010} (see Web Appendix (f) in that paper), by noting that
\begin{align*} 
\mbox{ED}_{k_l} = trace\left(\boldsymbol{Z}_k^{\top}\boldsymbol{P}\boldsymbol{Z}_k\boldsymbol{G}_k\frac{\boldsymbol{\Lambda}_{k_l}}{\sigma_{k_l}^{2}}\boldsymbol{G}_k\right) = trace\left(\boldsymbol{Z}_k\boldsymbol{G}_k\frac{\boldsymbol{\Lambda}_{k_l}}{\sigma_{k_l}^{2}}\boldsymbol{G}_k\boldsymbol{Z}_k^{\top}\left[\left(\boldsymbol{I}_n - \boldsymbol{P}_{\boldsymbol{X}}\right)\boldsymbol{V}\left(\boldsymbol{I}_n - \boldsymbol{P}_{\boldsymbol{X}}\right)\right]^{+}\right),
\end{align*}
where $\boldsymbol{\Gamma}^{+}$ denotes the Moore-Penrose pseudoinverse of $\boldsymbol{\Gamma}$. This equivalence has been proved, in a less general situation, by \cite{MX18} (see Web Appendix D in that paper). Following a similar reasoning, we obtain the upper bound for the effective degrees of freedom of the $k$-th random component
\[
\mbox{ED}_{k} \leq \mbox{rank}\left(\boldsymbol{X}, \boldsymbol{Z}_k\right) - \mbox{rank}\left(\boldsymbol{X}\right) = \mbox{rank}\left((I_n - \boldsymbol{P}_{\boldsymbol{X}})\boldsymbol{Z}_k\right) = \mbox{ED}^{ub}_{k}.
\]
Using the well known result that the rank of a matrix sum cannot exceed the sum of the ranks of the summand matrices, we have that $\mbox{ED}^{ub}_{k} \leq \sum_{l=1}^{p_k}\mbox{ED}^{ub}_{k_l}$. In general, however, this is a strict inequality, since most of the cases
\[
\mathcal{C}\left(\boldsymbol{Z}_k\boldsymbol{G}_k\boldsymbol{\Lambda}_{k_u}\right) \cap \mathcal{C}\left(\boldsymbol{Z}_k\boldsymbol{G}_k\boldsymbol{\Lambda}_{k_v}\right) \neq \left\{\boldsymbol{0}\right\}\;\;\;(u \neq v),
\]
where $\mathcal{C}\left(\boldsymbol{A}\right)$ denotes the linear space spanned by the columns of $\boldsymbol{A}$ \cite[see, e.g., Theorem 18.5.7 in][]{Harville1997}. Intuitively, we can interpret this as a sort of competition among the $p_k$ ``elements'' associated with the $k$-th random component. The $\mbox{ED}_{k_l}$ cannot vary ``free'' between $0$ and $\mbox{ED}^{ub}_{k_l}$, but they have to fulfil that their sum does not excess $\mbox{ED}^{ub}_{k}$.
\subsection{Computational aspects}\label{computaspects}
From a computational point of view, the evaluation of the expression given in eqn. (\ref{ed_per_component}) can be very costly. However, this computation can be relaxed by using the result given in (\ref{j_t_result}). For our purpose, it is easy to show that 
\begin{equation*}
\boldsymbol{G}_k\boldsymbol{Z}_k^{\top}\boldsymbol{P}\boldsymbol{Z}_k\boldsymbol{G}_k\boldsymbol{\Lambda}_{k_l} = \left(\boldsymbol{G}_k - \boldsymbol{C}^{*}_{kk}\right)\boldsymbol{\Lambda}_{k_l}.
\label{equivalence_hat_matrices} 
\end{equation*}
We note that $\boldsymbol{C}^{*}$ (i.e., the inverse of $\boldsymbol{C}$ in (\ref{MX:linearsystem_1})) is computed in order to estimate $\widehat{\boldsymbol{\beta}}$ and $\widehat{\boldsymbol{\alpha}}$ (see Appendix \ref{algorithm}). In addition, in those cases where $\boldsymbol{\Lambda}_{k_l}$ is diagonal, only the diagonal elements of $\left(\boldsymbol{G}_k - \boldsymbol{C}^{*}_{kk}\right)$ need to be explicitly obtained. This will considerably reduce the number of operations required, and therefore the computing time.

\section{Penalised smoothing and the SOP method}\label{OP_WWW}
This Section discusses several situations in the P-spline framework where estimation can be approached using the SOP method. As it will be seen, the method can be used whenever there are multiple penalties acting on the same coefficients. Anisotropic tensor-product P-splines is an example of overlapping penalties, and it has been extensively discussed in the paper by \cite{MXRA2015}. However, multiple penalties arise in a broader class of situations. We describe here two of those: Spatially-adaptive P-splines and P-splines for hierarchical curve data.
\subsection{Spatially-adaptive P-splines}\label{OP_Adaptive}
Consider a regression problem
\begin{equation}
y_i = f\left(x_i\right) + \epsilon_i \;\;\;\;\; i = 1,\ldots n,
\label{MX:PsplineM}
\end{equation}
where $f$ is a smooth and unknown function and $\epsilon_i \sim N\left(0, \phi\right)$. In the P-spline framework \citep{Eilers1996}, the unknown function $f(x)$ is approximated by a linear combination of $d$ B-splines basis functions, i.e., $f(x) = \sum_{j=1}^{d}\theta_{j}B_{j}\left(x\right)$. In matrix notation, model (\ref{MX:PsplineM}) is thus expressed as
\begin{equation}
\boldsymbol{y} = \boldsymbol{B}\boldsymbol{\theta} + \boldsymbol{\varepsilon},\;\;\boldsymbol{\varepsilon} \sim N\left(\boldsymbol{0}, \phi\boldsymbol{I}_n\right),
\label{MX:PsplineMM}
\end{equation} 
where $\boldsymbol{\theta} = \left(\theta_1, \theta_2, \ldots, \theta_d\right)^{\top}$ and $\boldsymbol{B}$ is a B-spline regression matrix of dimension $n \times d$, i.e., $b_{ij} = B_{j}\left(x_i\right)$ is the $j$-th B-spline evaluated at $x_i$. Smoothness is achieved by imposing a penalty on the regression coefficients $\boldsymbol{\theta}$ in the form 
\begin{equation}
\lambda\sum_{k = q + 1}^{d}\left(\Delta^q\theta_k\right)^2 = \lambda \boldsymbol{\theta}^{\top}\boldsymbol{D}_q^{\top}\boldsymbol{D}_q\boldsymbol{\theta},
\label{MX:1DPenalty}
\end{equation}
where $\lambda$ is the smoothing parameter, and $\Delta^q$ forms differences of order $q$ on adjacent coefficients, i.e., $\Delta\theta_k = \theta_k - \theta_{k-1}$, $\Delta^2\theta_k = \Delta\left(\Delta\theta_k\right) = \theta_k - \theta_{k-1} - \left(\theta_{k-1} - \theta_{k-2} \right) = \theta_k - 2\theta_{k-1} + \theta_{k-2}$, and so on for higher $q$. Finally, $\boldsymbol{D}_q$ is simply the matrix representation of $\Delta^q$.

\begin{figure}
 \begin{center}
 \includegraphics[width=12cm]{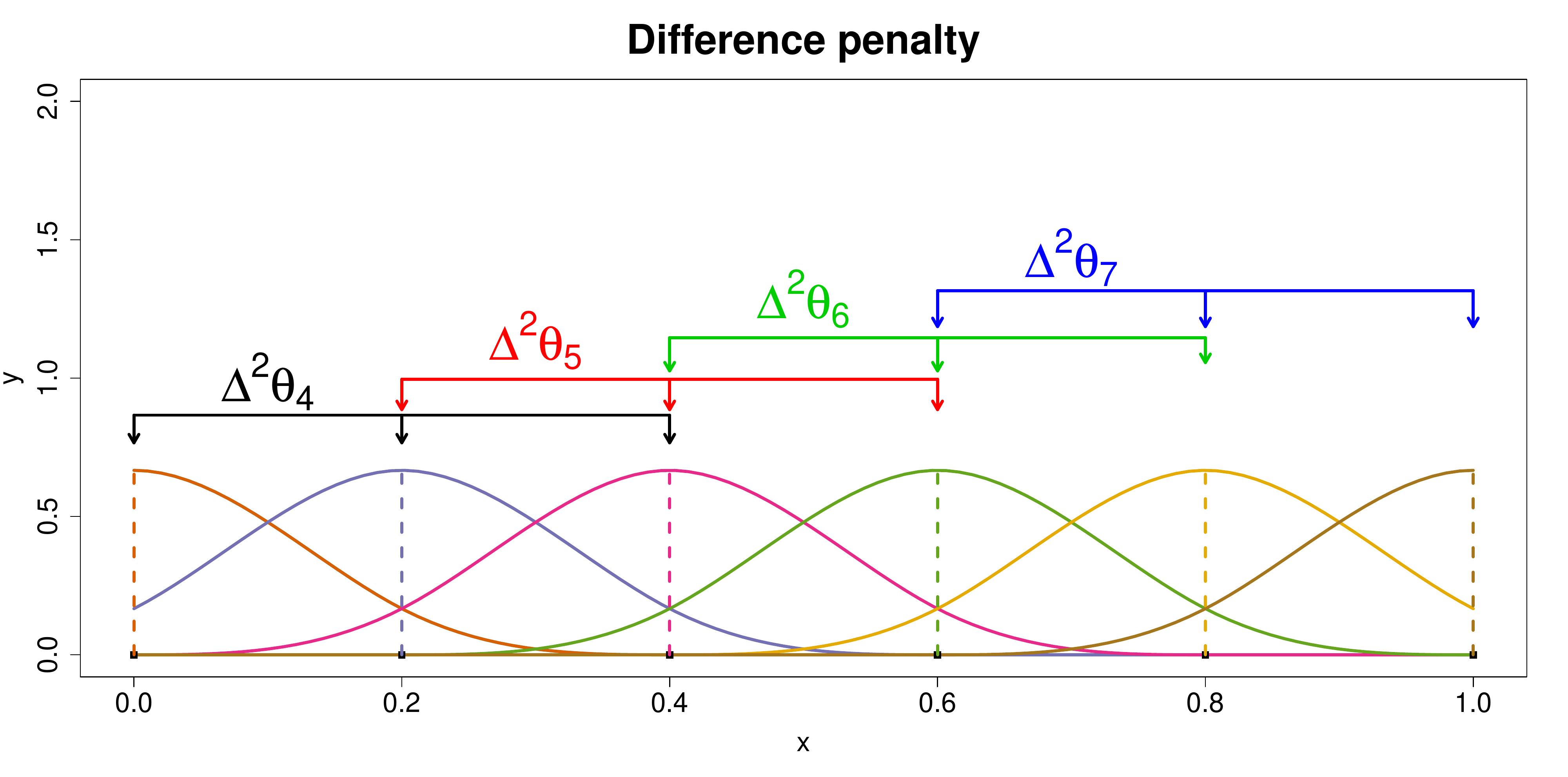}
	\end{center}
	\caption{Graphical representation of differences of order 2 on adjacent coefficients of cubic B-splines basis functions. Note the local and ordered nature of these differences.}
	\label{diff_penalty_1d}
\end{figure}

As can be observed in eqn.~(\ref{MX:1DPenalty}), the same smoothing parameter $\lambda$ applies to all coefficient differences, irrespective of their location (see Figure~\ref{diff_penalty_1d}). Thus, the model assumes that the same amount of smoothing is needed across the whole domain of the covariate. Adaptive P-splines \citep[see, e.g., ][among others]{Krivobokova08, Ruppert00} relax this assumption. The idea is simple, to replace the global smoothing parameter by smoothing parameters that vary locally according to the covariate value. This can be accomplished by specifying a different smoothing parameter for each coefficient difference \citep{Ruppert00, Wood2011}
\begin{equation}
\sum_{k = q + 1}^{d}\lambda_{k-q}\left(\Delta^q\theta_k\right)^2 = \boldsymbol{\theta}^{\top}\boldsymbol{D}_q^{\top}\mbox{diag}(\boldsymbol{\lambda})\boldsymbol{D}_q\boldsymbol{\theta},
\label{MX:AdaptPenalty}
\end{equation}
where $\boldsymbol{\lambda} = \left(\lambda_1,\ldots,\lambda_{d-q}\right)^{\top}$. Note that this approach would imply as many smoothing parameters as coefficient differences (i.e., $d - q$), which could lead to under-smoothing and unstable computations. Given the local and ordered nature of the coefficient differences (see Figure~\ref{diff_penalty_1d}), we may model the smoothing parameters $\lambda_k$ as a smooth function of $k$ (its position) and use B-splines for this purpose (here no penalty is assumed)
\begin{equation}
\boldsymbol{\lambda} = \boldsymbol{\Psi}\boldsymbol{\xi},
\label{MX:Adapt_smooth} 
\end{equation}
where $\boldsymbol{\Psi}$ is a B-spline regression matrix of dimension $(d-q)\times p$ with $p < (d-q)$, and $\boldsymbol{\xi} = \left(\xi_1,\ldots,\xi_{p}\right)^{\top}$ is the new vector of smoothing parameters. Performing some simple algebraic operations, it can be shown that the \textit{adaptive} penalty~(\ref{MX:AdaptPenalty}) is
\begin{equation}
\boldsymbol{\theta}^{\top}\left(\sum_{l=1}^{p}\xi_{l}\boldsymbol{D}_q^{\top} \mbox{diag}\left(\boldsymbol{\psi}_{l}\right)\boldsymbol{D}_q\right)\boldsymbol{\theta},
\label{MX:AdaptPenalty_smooth}
\end{equation}
where $\boldsymbol{\psi}_{l}$ denotes the column $l$ of $\boldsymbol{\Psi}$. Note that under this adaptive penalty, all coefficients are penalised by multiple smoothing parameters, i.e., there are overlapping penalties.

\subsubsection{Mixed model reparametrisation}
Estimation of the P-spline model (\ref{MX:PsplineMM}) subject to the adaptive penalty defined in~(\ref{MX:AdaptPenalty_smooth}) can be carried out based on the connection between P-splines and mixed models \citep[e.g., ][]{Currie2002, Wand2003}. It is easy to show that the null space (i.e., the unpenalised function space) of the adaptive penalty matrix $\boldsymbol{P}_{Ad} = \sum_{l=1}^{p}\xi_{l}\boldsymbol{D}_q^{\top} \mbox{diag}\left(\boldsymbol{\psi}_{l}\right)\boldsymbol{D}_q$ is independent of $\boldsymbol{\xi}$. In addition, note that when $\xi_u = \xi_v = \lambda$ $\left(\forall\;u, v\right)$ then
\[
\boldsymbol{P}_{Ad} = \sum_{l=1}^{p}\xi_{l}\boldsymbol{D}_q^{\top} \mbox{diag}\left(\boldsymbol{\psi}_{l}\right)\boldsymbol{D}_q = \lambda\boldsymbol{D}_q^{\top} \left(\sum_{l=1}^{p}\mbox{diag}\left(\boldsymbol{\psi}_{l}\right)\right)\boldsymbol{D}_q = \lambda\boldsymbol{D}_q^{\top}\boldsymbol{D}_q = \boldsymbol{P}.
\]
This is consequence of the rows of a B-spline regression matrix adding up to $1$. Thus, the null space of $\boldsymbol{P}_{Ad}$ is the same as that of $\boldsymbol{P}$. Different reparametrisations of P-spline models have been suggested in the literature \citep[see, e.g.,][]{Currie2002, Eilers1999}, all aiming to decompose the model into the unpenalised and the penalised part. The consequence of this decomposition is that the penalty matrix of the reparametrised P-spline model is of full rank, and so is the precision matrix of the corresponding mixed model. For our application, we use the proposal given in \cite{Eilers1999}. As will be seen, this approach gives rise to diagonal precision matrices. As discussed in Section \ref{computaspects}, this is very convenient from a computational point of view. Using \cite{Eilers1999}'s transformation, model (\ref{MX:PsplineMM}) is re-expressed as
\begin{equation*}
\boldsymbol{y} = \boldsymbol{B}\boldsymbol{\theta} + \boldsymbol{\varepsilon}= \boldsymbol{X}\boldsymbol{\beta} + \boldsymbol{Z}\boldsymbol{\alpha} + \boldsymbol{\varepsilon},
\end{equation*}
where $\boldsymbol{X}= \left[\boldsymbol{1}_n|\boldsymbol{x}|\ldots|\boldsymbol{x}^{(q-1)}\right]$ and $\boldsymbol{Z} = \boldsymbol{B}\boldsymbol{D}_q^{\top}\left(\boldsymbol{D}_q\boldsymbol{D}_q^{\top}\right)^{-1}$, and the precision (penalty) matrix of the vector of random (penalised) coefficients $\boldsymbol{\alpha}$ becomes
\begin{equation}
\boldsymbol{G}^{-1} = \frac{1}{\phi}\boldsymbol{F}^{\top}\boldsymbol{P}_{Ad}\boldsymbol{F} = \sum_{l=1}^{p}\sigma_l^{-2}\mbox{diag}\left(\boldsymbol{\psi}_{l}\right) = \sum_{l=1}^{p}\sigma_l^{-2}\widetilde{\boldsymbol{\Lambda}}_l,
\label{MX:Adaptive_G}
\end{equation}
where $\boldsymbol{F} = \boldsymbol{D}_q^{\top}\left(\boldsymbol{D}_q\boldsymbol{D}_q^{\top}\right)^{-1}$, $\widetilde{\boldsymbol{\Lambda}}_l = \mbox{diag}\left(\boldsymbol{\psi}_{l}\right)$, and $\sigma_l^2 = \phi/\xi_l$. Thus, the precision matrix is linear in the precision parameters $\sigma_l^{-2}$, and the SOP method can therefore be used. 

We note that the model has a single random component ($c = 1$). The reparametrisation ensures that $\mbox{rank}(\boldsymbol{X}, \boldsymbol{Z}) = \mbox{rank}(\boldsymbol{X}) + \mbox{rank}(\boldsymbol{Z})$. Thus, $\mbox{rank}(\boldsymbol{X}, \boldsymbol{Z}\widetilde{\boldsymbol{\Lambda}}_l) = \mbox{rank}(\boldsymbol{X}) + \mbox{rank}(\boldsymbol{Z}\widetilde{\boldsymbol{\Lambda}}_l) > \mbox{rank}(\boldsymbol{X})$. Exploiting the fact that $\boldsymbol{G}$ and $\widetilde{\boldsymbol{\Lambda}}_l$ are diagonal, and thus commute, condition (i) of Theorem \ref{the2} is satisfied.
\subsection{P-splines for hierarchical curve data}\label{OP_SSC}
For simplicity, let's assume balanced hierarchical curve data. Our data consists on $m$ individuals each with $s$ different measurements at times $\boldsymbol{t} = (t_1, t_2,\ldots,t_s)$. Our interest is focused in
\begin{equation}
y_{ij} = f\left(t_i\right) + g_j\left(t_i\right) + \varepsilon_{ij} \;\; 1\leq i \leq s,\;1\leq j \leq m,
\label{SSC_model_ind}
\end{equation}
where $y_{ij}$ is the response variable on the $j$-th subject at time $t_{i}$, $f$ is a function describing the population effect, $g_j$ are random functions measuring the deviation of the $j$-th subject from the population effect, and $\varepsilon_{ij} \sim N\left(0, \phi\right)$. A simple model would consist in a parametric specification for $f$ and $g_j$, e.g., $f\left(t\right) = \beta_0 + \beta_1t$ and $g_i\left(t\right) = \alpha_{0j} + \alpha_{1j}t$, with $\boldsymbol{\alpha}_0 = \left(\alpha_{01}, \ldots, \alpha_{0m}\right)^{\top} \sim N\left(\boldsymbol{0}, \sigma_{0}^2\boldsymbol{I}_m\right)$ and $\boldsymbol{\alpha}_1 = \left(\alpha_{11}, \ldots, \alpha_{1m}\right)^{\top} \sim N\left(\boldsymbol{0}, \sigma_{1}^2\boldsymbol{I}_m\right)$. 

A more flexible approach consists in assuming $f$ and $g_j$ smooth and unknown. Important contributions in the P-spline framework can be found in \cite{Durban05} and \cite{Ruppert2003}. Both approaches are based on modelling $f$ and $g_j$ by means of truncated line basis, and estimation is based on penalised methods and linear mixed model techniques. More recently, \cite{Djeundje10} extend those models, and propose the inclusion of an extra penalty for the individual curve coefficients. The authors argue that this extra penalty is needed to address identifiability issues when estimating the population effect. Under the so-called \texttt{M0} model in \cite{Djeundje10}'s paper, model (\ref{SSC_model_ind}) is expressed in matrix notation as
\[
\boldsymbol{y}_{\cdot j} = \underbrace{\boldsymbol{B}\boldsymbol{\theta}}_{f\left(\boldsymbol{t}\right)} + \underbrace{\breve{\boldsymbol{B}}\breve{\boldsymbol{\theta}}_j}_{g_j\left(\boldsymbol{t}\right)} + \boldsymbol{\varepsilon}_{\cdot j},
\]
where $\boldsymbol{y}_{\cdot j} = \left(y_{1j},\ldots, y_{sj}\right)^{\top}$ is the response vector for the $j$-th individual (the same holds for $\boldsymbol{\varepsilon}_{\cdot j}$), and $\boldsymbol{B}$ and $\breve{\boldsymbol{B}}$ are B-spline regression matrices of possibly different size for, respectively, the population curve and the individual deviation. The vector $\boldsymbol{\theta}$ is assumed fixed, but subject to a $q$-th order penalty of the form $\boldsymbol{P} = \lambda_1\boldsymbol{D}_q^{\top}\boldsymbol{D}_q$, and $\breve{\boldsymbol{\theta}}_j$ is a random vector with distribution $N\left(\boldsymbol{0}, \phi\breve{\boldsymbol{P}}^{-1}\right)$ where 
\begin{equation}
\breve{\boldsymbol{P}} = \lambda_2\breve{\boldsymbol{D}}_{\breve{q}}^{\top}\breve{\boldsymbol{D}}_{\breve{q}} + \lambda_3\boldsymbol{I}_{\breve{d}}.
\label{SSC_P_ind}
\end{equation}
The first term $\lambda_2\breve{\boldsymbol{D}}_{\breve{q}}^{\top}\breve{\boldsymbol{D}}_{\breve{q}}$ is responsible for the smoothness of the individuals curves, whereas $\lambda_3\boldsymbol{I}_{\breve{d}}$ addresses the identifiability aspect \citep[see][for more details]{Djeundje10}. Note that each random effect is shrunk (penalised) by both smoothing parameters $\lambda_2$ and $\lambda_3$, and thus the precision matrix $\breve{\boldsymbol{G}}^{-1} = 1/\phi\breve{\boldsymbol{P}}$ is linear in the precision parameters
\begin{equation}
\breve{\boldsymbol{G}}^{-1} = \sum_{l = 2}^{3}\sigma_l^{-2}\breve{\boldsymbol{\Lambda}}_l,
\label{SSC_G_ind}
\end{equation}
where $\sigma_2^2 = \phi/\lambda_2$ and $\sigma_3^2 = \phi/\lambda_3$ and $\breve{\boldsymbol{\Lambda}}_2 = \breve{\boldsymbol{D}}_{\breve{q}}^{\top}\breve{\boldsymbol{D}}_{\breve{q}}$ and $\breve{\boldsymbol{\Lambda}}_3 = \boldsymbol{I}_{\breve{d}}$. 

In more compact way, we express the model for the whole sample as
\begin{equation}
\boldsymbol{y} = [\boldsymbol{1}_{m}\otimes\boldsymbol{B}]\boldsymbol{\theta} + [\boldsymbol{I}_{m}\otimes\breve{\boldsymbol{B}}]\breve{\boldsymbol{\theta}} + \boldsymbol{\varepsilon},
\label{SSC_model_pop}
\end{equation}
where $\otimes$ denote the Kronecker product, $\boldsymbol{y} = \left(\boldsymbol{y}_{\cdot 1}^{\top},\ldots, \boldsymbol{y}_{\cdot m}^{\top}\right)^{\top}$, $\breve{\boldsymbol{\theta}} = \left(\breve{\boldsymbol{\theta}}_{1}^{\top},\ldots, \breve{\boldsymbol{\theta}}_{m}^{\top}\right)^{\top}$, $\boldsymbol{\varepsilon} = \left(\boldsymbol{\varepsilon}_{\cdot,1}^{\top},\ldots, \boldsymbol{\varepsilon}_{\cdot,m}^{\top}\right)^{\top}$ and
\[
\breve{\boldsymbol{\theta}} \sim N\left(\boldsymbol{0}, \boldsymbol{I}_{m}\otimes\breve{\boldsymbol{G}}\right).
\]
One last step is needed in order to apply the SOP method for the estimation of model (\ref{SSC_model_pop}): the decomposition of the population effect into the unpenalised and the penalised part. We use here the approach based on the eigenvalue decomposition (EVD) of the penalty. Let $\boldsymbol{D}^{\top}_q\boldsymbol{D}_q = \boldsymbol{U}\boldsymbol{\Sigma}\boldsymbol{U}^{\top}$ be the EVD of $\boldsymbol{D}_q^{\top}\boldsymbol{D}_q$. Here $\boldsymbol{U}$ denotes the matrix of eigenvectors and $\boldsymbol{\Sigma}$ the diagonal matrix of eigenvalues. Let us also denote by $\boldsymbol{U}_{+}$ ($\boldsymbol{\Sigma}_{+}$) and $\boldsymbol{U}_{0}$ ($\boldsymbol{\Sigma}_{0}$) the sub-matrices corresponding to the non-zero and zero eigenvalues, respectively. In this case, model (\ref{SSC_model_pop}) is re-expressed as
\[
\boldsymbol{y} = \boldsymbol{X}\boldsymbol{\beta} + \boldsymbol{Z}\boldsymbol{\alpha} + \boldsymbol{\varepsilon},
\]
where $\boldsymbol{\beta} =\boldsymbol{U}_{+}^{\top}\boldsymbol{\theta}$, $\boldsymbol{\alpha} = \left(\left(\boldsymbol{U}_{0}^{\top}\boldsymbol{\theta}\right)^{\top}, \breve{\boldsymbol{\theta}}^{\top}\right)^{\top}$, $\boldsymbol{X} = [\boldsymbol{1}_m\otimes\boldsymbol{B}\boldsymbol{U}_{0}]$ and $\boldsymbol{Z} = [\boldsymbol{1}_m\otimes\boldsymbol{B}\boldsymbol{U}_{+}:\boldsymbol{I}_{m}\otimes\breve{\boldsymbol{B}}]$. Finally, $\boldsymbol{\alpha} \sim N\left(\boldsymbol{0}, \boldsymbol{G}\right)$, where
\[
\boldsymbol{G}^{-1} = 
\begin{pmatrix}
\frac{1}{\sigma_{1}^2}\boldsymbol{\Sigma}_ {+} & \boldsymbol{0}_{d\times\left(m \breve{d}\right)}\\
\boldsymbol{0}_{\left(m \breve{d}\right)\times c} & \boldsymbol{I}_{m}\otimes\breve{\boldsymbol{G}}^{-1}
\end{pmatrix} = \sum_{l=1}^{3}\sigma_l^{-2}\widetilde{\boldsymbol{\Lambda}}_{l},
\]
with $\sigma_{l}^2 = \frac{\phi}{\lambda_l}$ and 
\[
\widetilde{\boldsymbol{\Lambda}}_{1} = 
\begin{pmatrix}
\boldsymbol{\Sigma}_{+} & \boldsymbol{0}_{d\times\left(m \breve{d}\right)}\\
\boldsymbol{0}_{\left(m \breve{d}\right)\times d} & \boldsymbol{0}_{\left(m \breve{d}\right)\times\left(m \breve{d}\right)}
\end{pmatrix}
\qquad
\widetilde{\boldsymbol{\Lambda}}_{2} = 
\begin{pmatrix}
\boldsymbol{0}_{d \times d} & \boldsymbol{0}_{d\times\left(m \breve{d}\right)}\\
\boldsymbol{0}_{\left(m \breve{d}\right)\times d} & \boldsymbol{I}_{m} \otimes \breve{\boldsymbol{\Lambda}}_2
\end{pmatrix}
\qquad
\widetilde{\boldsymbol{\Lambda}}_{3} = 
\begin{pmatrix}
\boldsymbol{0}_{d \times d} & \boldsymbol{0}_{d \times\left(m \breve{d}\right)}\\
\boldsymbol{0}_{\left(m \breve{d}\right)\times d} & \boldsymbol{I}_{m} \otimes \breve{\boldsymbol{\Lambda}}_3
\end{pmatrix}.
\]
The precision matrix is linear in the precision parameters, and thus appropriate for the SOP method. There are, in this case, two random components, modelling, respectively, the population curve and the individual deviations, with 
\begin{align*}
\boldsymbol{Z}_1 & = \boldsymbol{1}_m\otimes\boldsymbol{B}\boldsymbol{U}_{+} & \mbox{and}\;\;\;\boldsymbol{G}^{-1}_1 & = \sigma_1^{-2}\boldsymbol{\Sigma}_{+},\\
\boldsymbol{Z}_2 & = \boldsymbol{I}_{m}\otimes\breve{\boldsymbol{B}} & \mbox{and}\;\;\;\boldsymbol{G}^{-1}_2 & = \sigma_2^{-2}\boldsymbol{I}_{m} \otimes \breve{\boldsymbol{\Lambda}}_2 + \sigma_3^{-2}\boldsymbol{I}_{m} \otimes \breve{\boldsymbol{\Lambda}}_3 = \sigma_{2}^{-2} \boldsymbol{I}_{m} \otimes \breve{\boldsymbol{D}}_{\breve{q}}^{\top}\breve{\boldsymbol{D}}_{\breve{q}} + \sigma_{3}^{-2}\boldsymbol{I}_{m}\otimes\boldsymbol{I}_{\breve{d}}\;.
\end{align*}
Recall that $\boldsymbol{X} = \boldsymbol{1}_m\otimes\boldsymbol{B}\boldsymbol{U}_{0}$, and note that $\mbox{rank}\left(\boldsymbol{1}_m\otimes\boldsymbol{B}\boldsymbol{U}_{0}\right) = \mbox{rank}\left(\boldsymbol{B}\boldsymbol{U}_{0}\right)$. We now show that condition (i) of Theorem \ref{the2} is satisfied for all variance parameters:
\begin{description}
\item[$\sigma_1^2$:] By construction, $\boldsymbol{\Sigma}_{+}$ is positive definite and of full rank, and $\mbox{rank}\left(\boldsymbol{X}, \boldsymbol{Z}_1\right) = \mbox{rank}\left(\boldsymbol{X}\right) + \mbox{rank}\left(\boldsymbol{Z}_1\right)$. Noting that  $\mbox{rank}\left(\boldsymbol{X}, \boldsymbol{Z}_1\boldsymbol{G}_1\boldsymbol{\Sigma}_{+}\right) = \mbox{rank}\left(\boldsymbol{X}, \boldsymbol{Z}_1\right)  > \mbox{rank}\left(\boldsymbol{X}\right)$, the condition is verified.
\item[$\sigma_2^2$:] By e.g. Corollary 18.2.2 in \cite{Harville1997}, $\boldsymbol{G}_2$ and $\boldsymbol{I}_{m} \otimes \breve{\boldsymbol{D}}_{\breve{q}}^{\top}\breve{\boldsymbol{D}}_{\breve{q}}$ commute. Thus, as long as $m > 1$, it is easy to show that
\[
\mbox{rank}\left(\boldsymbol{X}, \boldsymbol{Z}_2\boldsymbol{I}_{m} \otimes \breve{\boldsymbol{D}}_{\breve{q}}^{\top}\breve{\boldsymbol{D}}_{\breve{q}}\right) = \mbox{rank}\left(\boldsymbol{1}_m\otimes\boldsymbol{B}\boldsymbol{U}_{0}, \boldsymbol{I}_{m}\otimes\breve{\boldsymbol{B}}\breve{\boldsymbol{D}}_{\breve{q}}^{\top}\breve{\boldsymbol{D}}_{\breve{q}}\right) > \mbox{rank}\left(\boldsymbol{B}\boldsymbol{U}_{0}\right). 
\]
\item[$\sigma_3^2$:] Note that $\boldsymbol{G}_2$ and $\boldsymbol{I}_{m}\otimes\boldsymbol{I}_{\breve{d}}$ commute. As before, as long as $m > 1$, it is easy to show that
\[
\mbox{rank}\left(\boldsymbol{X}, \boldsymbol{Z}_2\boldsymbol{I}_{m}\otimes\boldsymbol{I}_{\breve{d}}\right) = \mbox{rank}\left(\boldsymbol{1}_m\otimes\boldsymbol{B}\boldsymbol{U}_{0}, \boldsymbol{I}_{m}\otimes\breve{\boldsymbol{B}}\right) > \mbox{rank}\left(\boldsymbol{B}\boldsymbol{U}_{0}\right). 
\]
\end{description}
\section{Examples}\label{examples}
This Section presents several data examples where the SOP method represents a powerful alternative to existing estimation procedures. We discuss three different analyses: the first two examples are concerned with spatially-adaptive P-splines, but each of them deals with a different situation regarding complexity and aim; the last example is devoted to illustrating our method for the analysis of hierarchical curve data. All computations were performed in (64-bit) \texttt{R} 3.4.4 \citep{R16}, and a 2.30GHz $\times$ 4 Intel$^\circledR$ Core$\texttrademark$ i5 processor computer with 15.6GB of RAM and Ubuntu 16.04 LTS operating system.
\subsection{Doppler function}\label{doppler_example}
\begin{figure}
 \begin{center}
 \subfigure[True function]{\includegraphics[width=6cm]{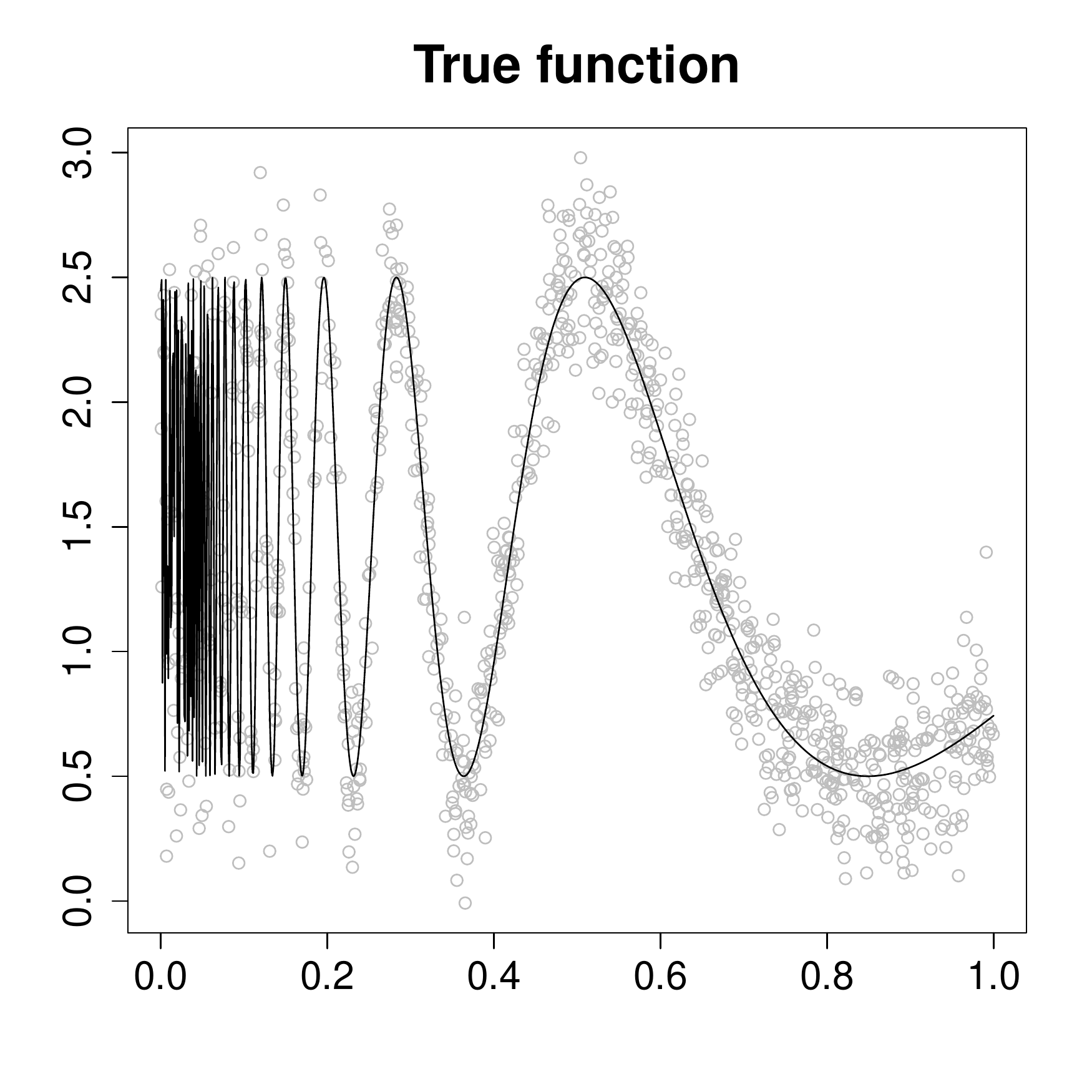}\label{MX_df_true}}
 \subfigure[SOP method without adaptive penalty]{\includegraphics[width=6cm]{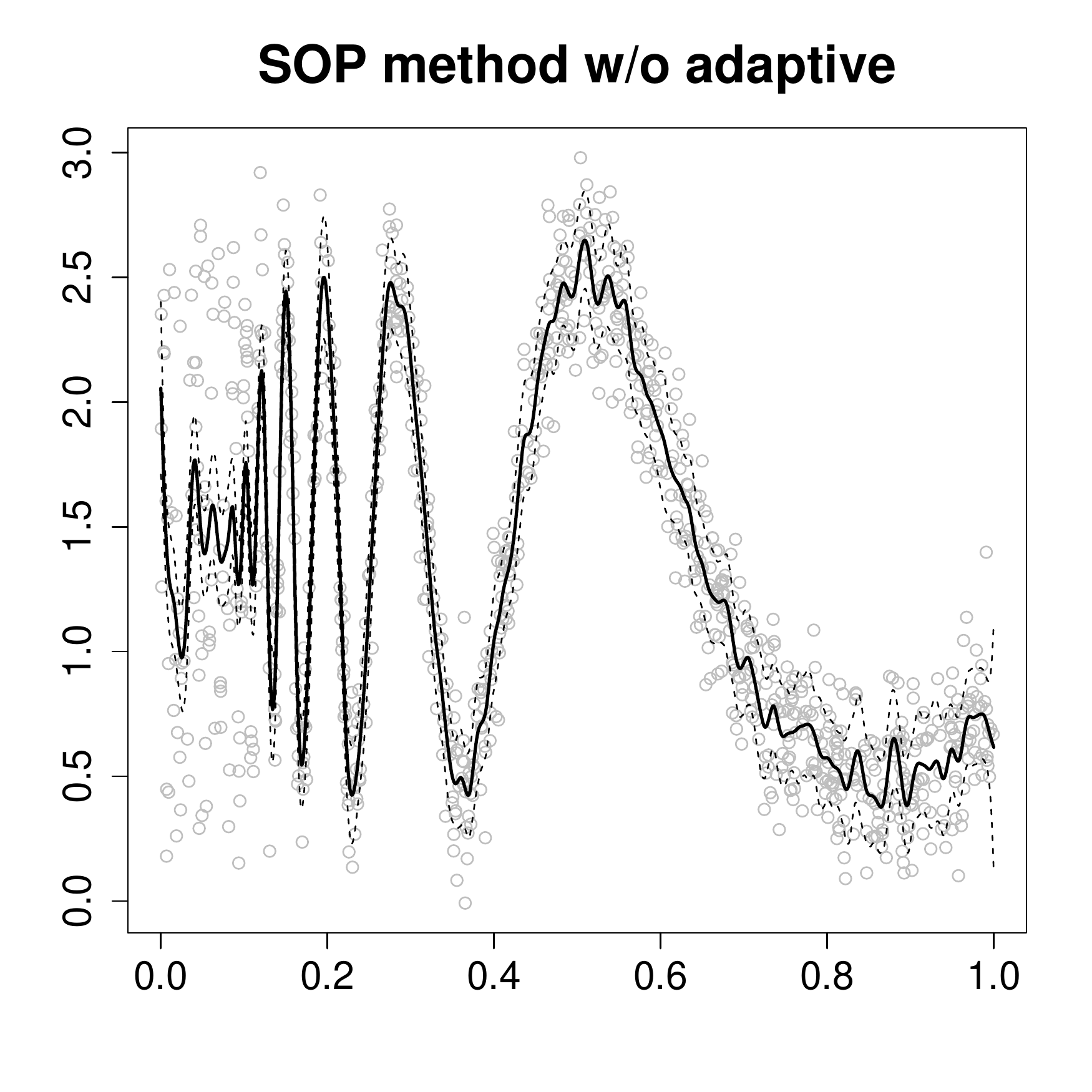}\label{MX_df_SOP_na}}
	 \subfigure[SOP method with adaptive penalty]{\includegraphics[width=6cm]{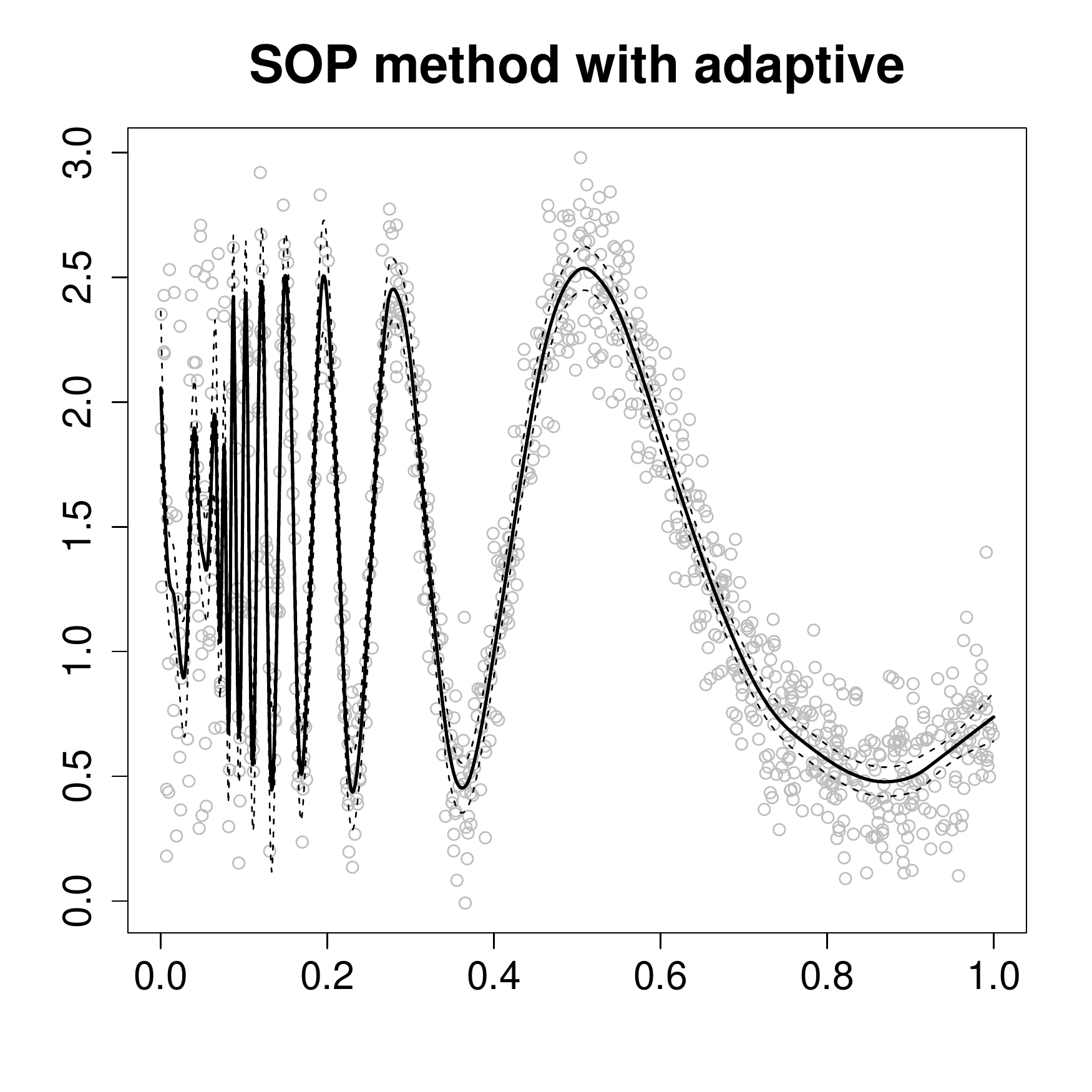}\label{MX_df_SOP}}
	 \subfigure[\texttt{mgcv} package with adaptive penalty]{\includegraphics[width=6cm]{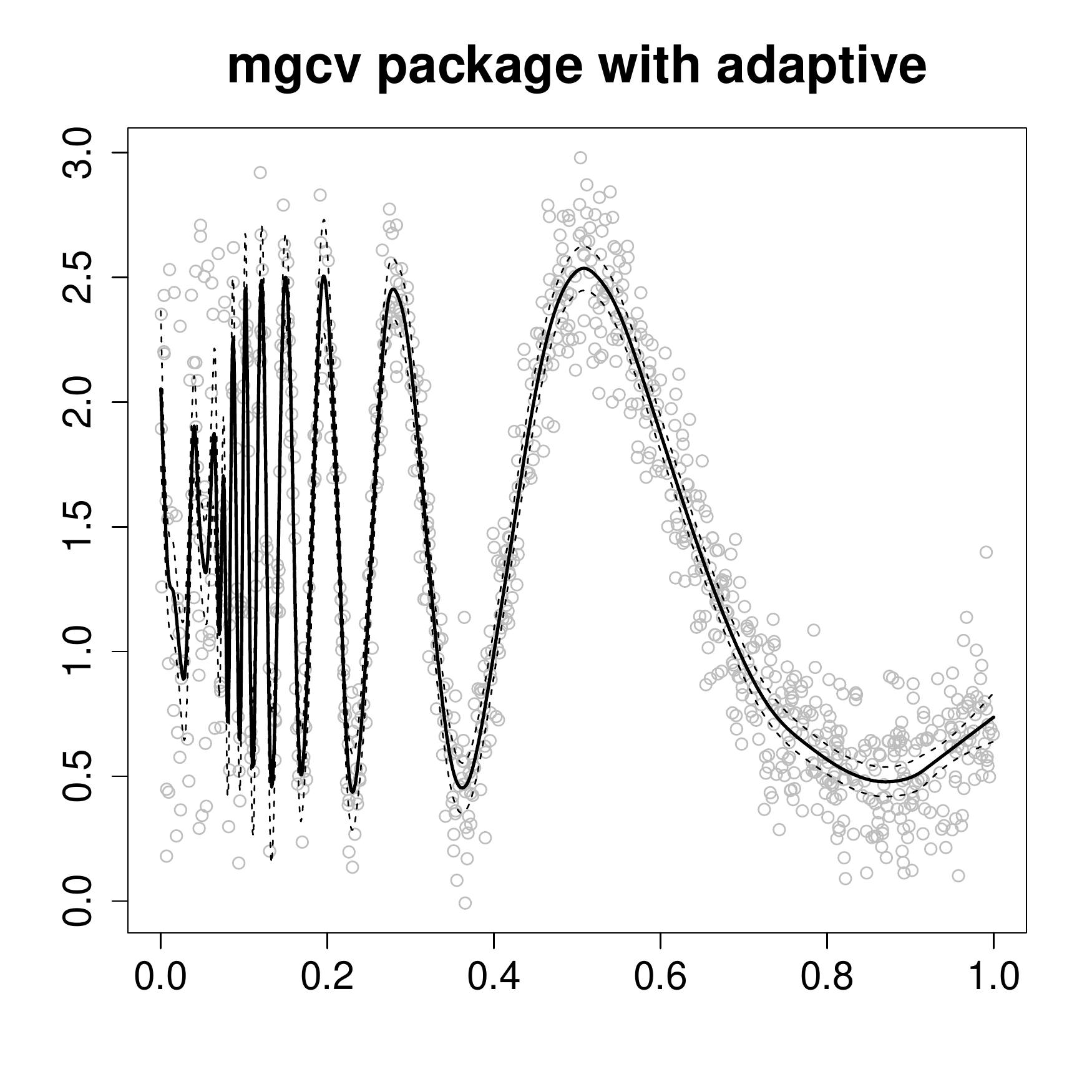}\label{MX_df_mgcv}}
	\end{center}
		 \caption{For the Doppler function: (a) True function (solid line) and simulated data points (grey points), (b) estimated curve using the SOP method without adaptive penalty (solid line) jointly with $95\%$ approximate confidence intervals (dotted lines), (c) estimated curve using the SOP method and adaptive penalty (solid line) jointly with $95\%$ approximate confidence intervals (dotted lines); and (d) estimated curve using the \texttt{mgcv} package (solid line) jointly with $95\%$ approximate confidence intervals (dotted lines).}
		 \label{Doppler_function_results}
\end{figure}

For our first example, we consider the Doppler function. This is a common example in the adaptive smoothing literature, and has been discussed, by \cite{Ruppert00, Krivobokova08, Tibshirani14}, among others. Data are generated according to
\[
y_i = \sin\left(4/x_i\right) + 1.5 + \varepsilon_i, \;\;\;i = 1,\ldots, n,
\]
where $x_i \sim U\left[0,1\right]$, $\varepsilon_i \sim N\left(0, 0.2^2\right)$, and $n = 1000$. FFor fitting the data, we assume the spatially-adaptive P-spline model discussed in Section \ref{OP_Adaptive}. We compare the performance of the SOP method with that implemented in the \texttt{R}-package \texttt{mgcv}, version \texttt{1.8-23}, and described in \cite{Wood2011}. It is worth noticing that both approaches implement in essence the same adaptive P-spline model; the only difference is the estimation procedure (and, possibly, the reparametrisation). In addition, we also fit the model without assuming an adaptive penalty. In this case, the SOP method reduces to Harville's approach (see Section \ref{HfD}). In all cases, we use $200$ cubic B-splines to represent the smooth function, jointly with second-order differences. For the adaptive approaches, $15$ equally-spaced cubic B-splines are used for the smoothing parameters (see (\ref{MX:Adapt_smooth})). These values are chosen to provide enough flexibility to the model. Under this configuration, there are a total of $15$ variance parameters. Figure \ref{MX_df_true} shows the true simulated Doppler function. Figures \ref{MX_df_SOP_na} and \ref{MX_df_SOP} show, respectively, the estimated curves based on the SOP method without and with an adaptive penalty. Results using \texttt{mgcv} are depicted in Figure \ref{MX_df_mgcv}. As expected, both adaptive approaches perform similarly. With the specified configuration, they are able to capture $7$ cycles of the Doppler function. On the other hand, the non-adaptive approach is able to capture only $4$ cycles, and presents very wiggly estimates, especially on the right-hand side of the covariate domain. In terms of EDs, for the SOP model without adaptive penalty, we obtain a total ED of $95.8$ (out of $200$). For models with adaptive smoothing we obtained identical results, i.e. $50.2$ (with SOP) and $50.0$ (with \texttt{mgcv}). It is worth remembering that, using the SOP method the total ED is obtained by adding up the EDs associated with each variance parameter in the model (plus the dimension of the fixed part). These EDs are the denominator of the estimate update expressions of the variance parameters. The gain of the SOP method is clear when we compare the computing times: $1.0$ second with our approach ($0.4$ if we do not consider adaptive), and $45$ seconds using \texttt{mgcv}.
\subsection{X-ray diffraction data}
\begin{figure}
 \begin{center}
 \includegraphics[width=15cm]{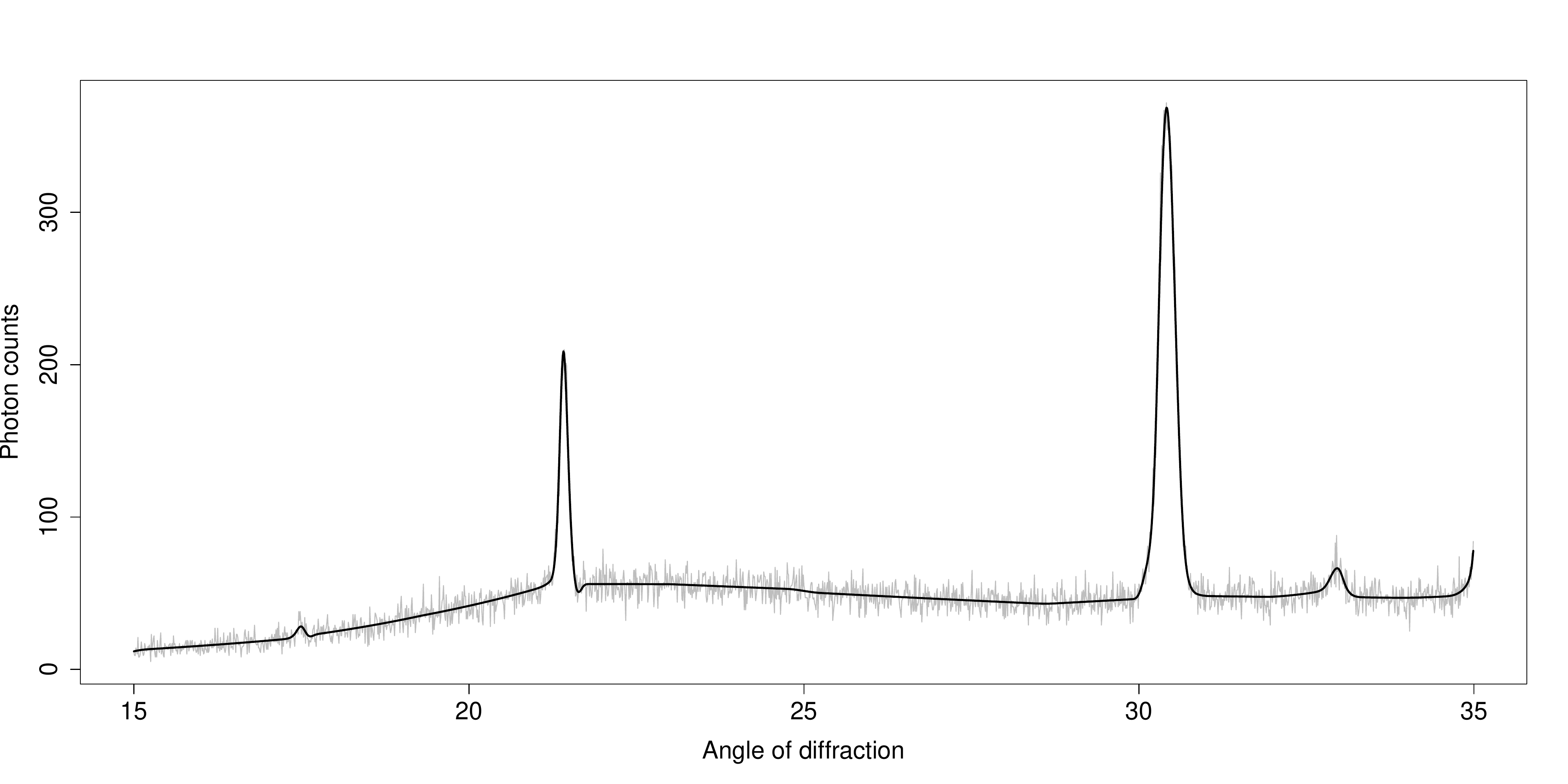}
	\end{center}
		 \caption{For the x-ray radiation example: estimated smooth effect of the angle of diffraction on the x-ray radiation using the SOP method (solid black line). The grey lines represent the raw data. }
	\label{mx_x_ray}
\end{figure}

For this example, we use data from a X-ray crystallography radiation scan of a thin layer of indium tin oxide. X-ray crystallography allows the exploration of the molecular and atomic structure of crystals. Crystallographers precisely rotate the crystal by entering its desired orientation while it is illuminated by a beam of X-rays. Depending on the angle, the number of diffracted photons varies and they are detected and counted by an electronic detector. The data set was analysed by \cite{Davies08, Davies13} and can be found in the R-package \texttt{diffractometry} as \texttt{indiumoxide}. Figure~\ref{mx_x_ray} shows such an X-ray diffraction scan (grey lines). The aim of X-ray diffraction analysis is to determine (a) the signal baseline (and remove it); and (b) the number of peaks (and isolate them to further analysis of their position, height, symmetry, and so forth). This example is solely included to illustrate the potentiality of the method presented in this paper for the analysis of very complex data. For a different modelling approach, see \cite{Camarda16}. Given that the outcome variable represents count data, a Poisson model is adopted
\[
E\left[y_i \mid x_i\right] = \exp\left(f\left(x_i\right)\right),
\]
where $y_i$ and $x_i$ $(i = 1, \ldots,2000)$ denote, respectively, the photon counts and the angle of diffraction. To provide enough flexibility to the model in order to make it able to capture the peaks (see Figure~\ref{mx_x_ray}), we use $200$ cubic B-splines and second-order differences for the function, and $80$ cubic B-splines for the adaptive penalty. Results are shown in Figure~\ref{mx_x_ray}. The results using the \texttt{mgcv} package are almost identical to our proposal, and are not depicted. In this case, our method takes less than $3$ seconds, whereas \texttt{mgcv} is around $750$ times slower ($33$ minutes). Regarding the EDs, we obtain a total of $32.1$ and $29.0$ using SOP and \texttt{mgcv}, respectively. 

\subsection{Diffusion tensor imaging scan data}\label{DTI_example}
\begin{figure}
 \begin{center}
 \subfigure[FA profiles]{\includegraphics[width=14cm]{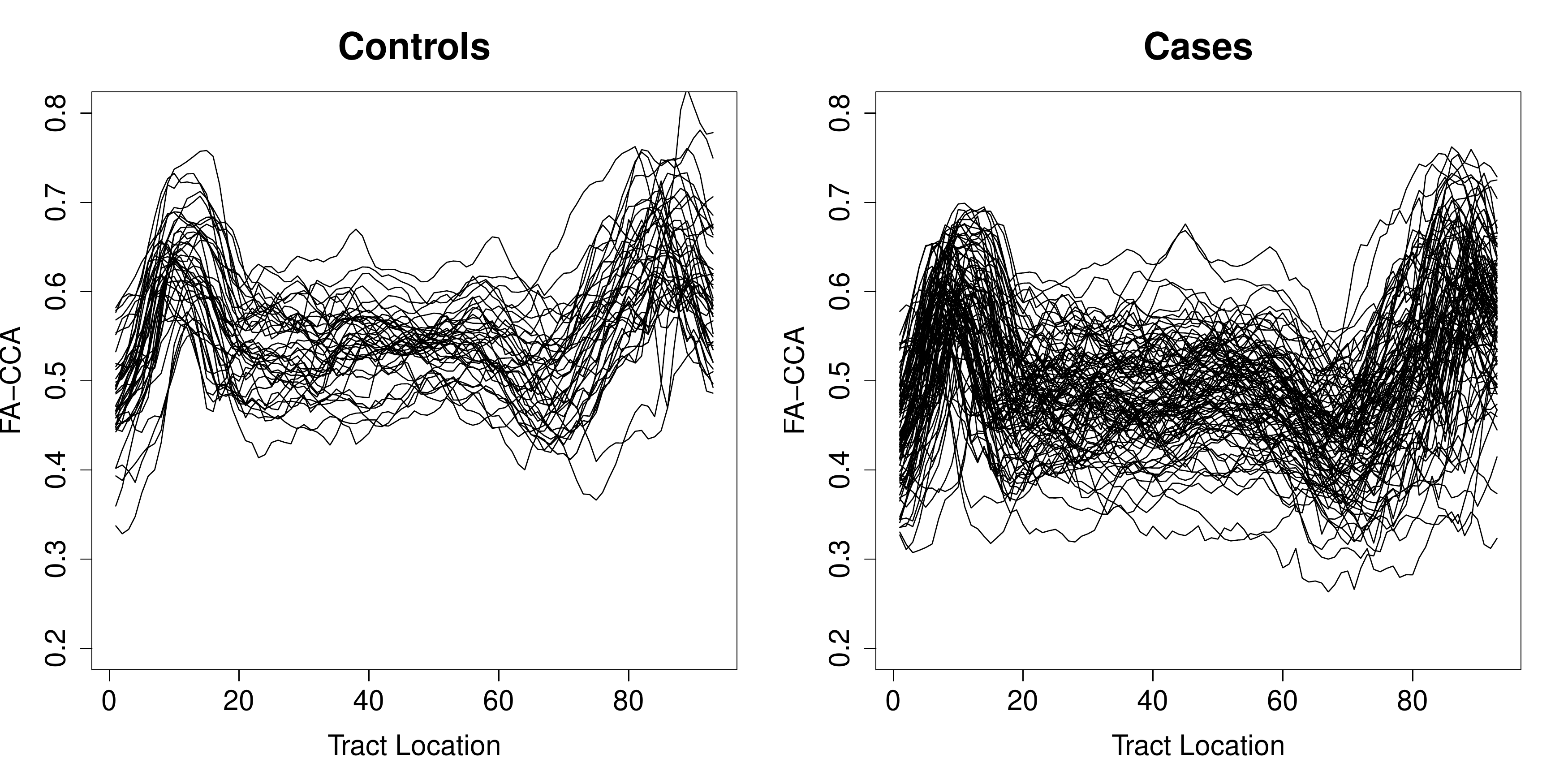}\label{DTI_raw}}
 \subfigure[Estimated FA profile for MS patiens]{\includegraphics[width=7cm]{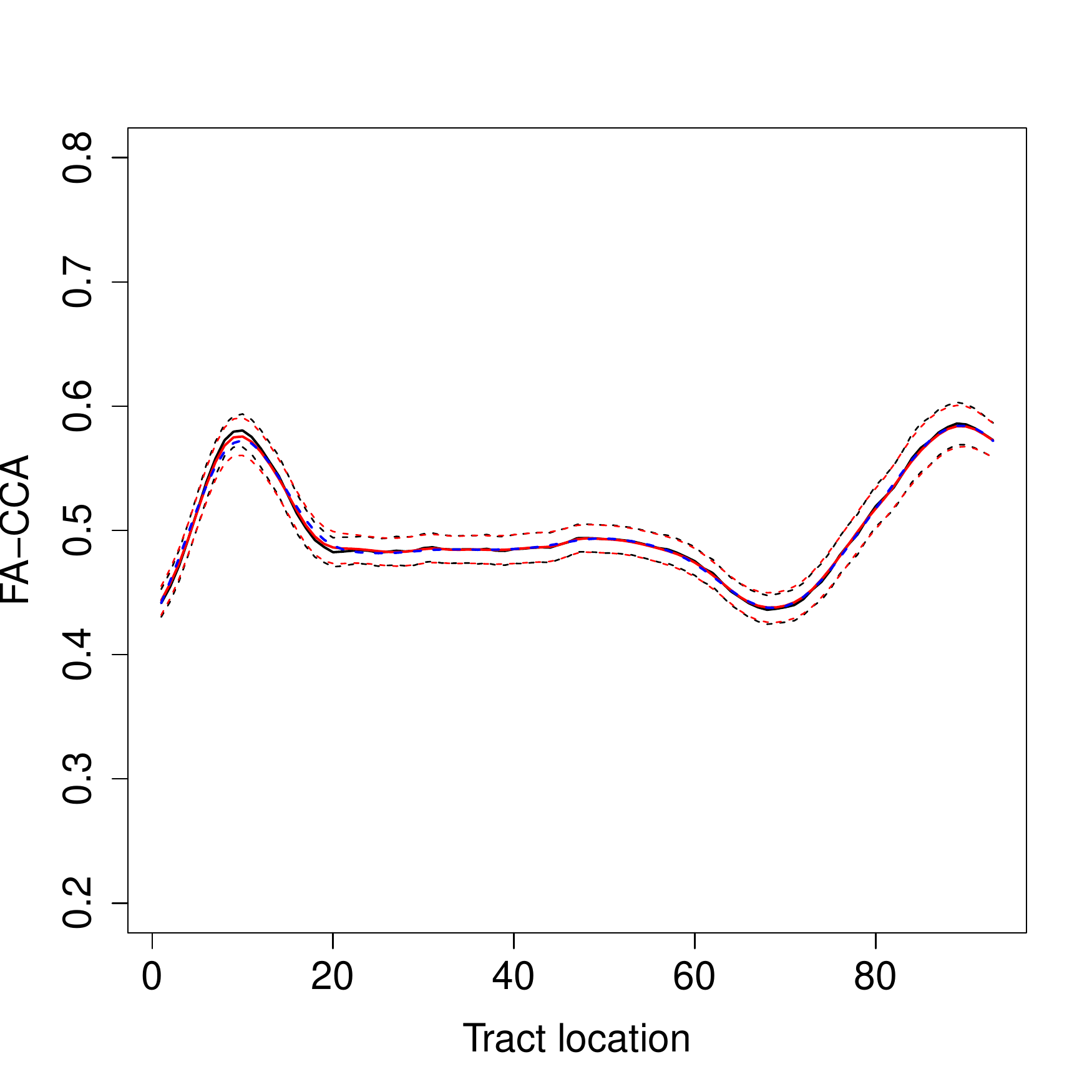}\label{DTI_results_pop}}
	 \subfigure[Individual FA profiles]{\includegraphics[width=7cm]{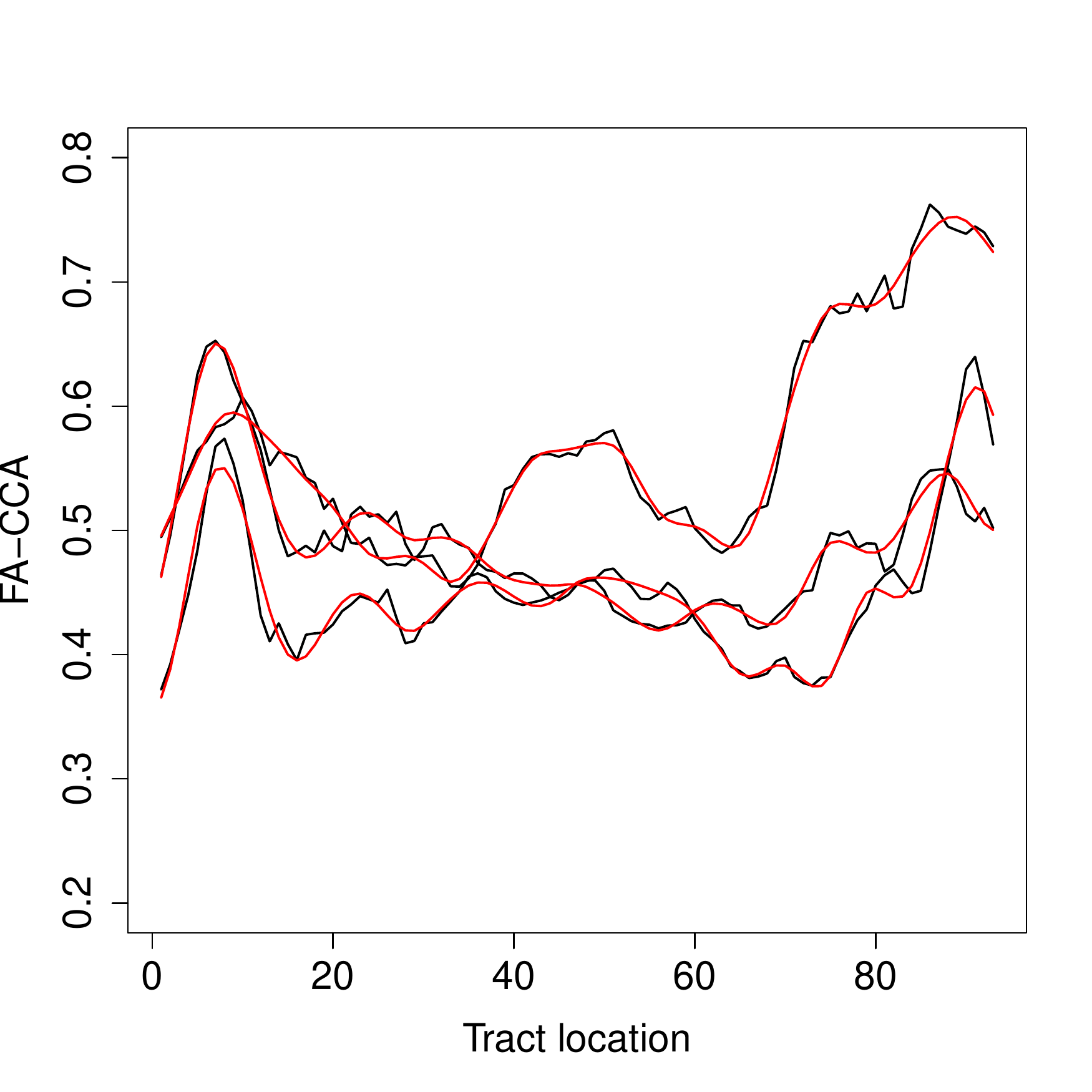}\label{DTI_results_ind}}
	\end{center}
		 \caption{For the diffusion tensor imaging scan data: (a) observed FA values along the CCA tract. Left: healthy controls. Right: MS patients; (b) estimated population (group) FA profile for individuals affected with MS. Solid red line: SOP method. Dotted blue line: \cite{Djeundje10}'s code. The dotted red lines are the pointwise $95\%$ confidence intervals based on the full-sandwich standard errors proposed by \cite{Heckman13}. The back line is the observed mean and the dotted black lines are the empirical confidence intervals; and (c) estimated (red lines) and observed (black lines) individual FA profiles for $3$ selected individuals.}
\end{figure}

Our last example deals with hierarchical curve data. We analyse the \texttt{DTI} dataset, that can be found in the \texttt{R}-package \texttt{refund} \citep{Goldsmith16}. A detailed description of the study and data can be found in \cite{Goldsmith11}, \cite{Goldsmith12} and \cite{Greven17}. In brief, the study aimed at comparing the white matter tracts in patients affected by multiple sclerosis (MS) and healthy individuals. Multiple sclerosis is a disease of the central nervous system that causes lesions in white matter tracts, thus interrupting the travel of nerve impulses to and from the brain and spinal cord. Diffusion tensor imaging (DTI) is a magnetic resonance imaging technique which makes it possible to study the white matter tracts by tracing the diffusion of water on the brain. From DTI scans, fractional anisotropy (FA) measurements can be extracted. FA is related to water diffusion, and thus to MS diagnosis and progression. The \texttt{DTI} dataset contains FA measurements of healthy and diseased individuals, recorded at several locations along the callosal fiber tract (CCA) and the right corticospinal tract in the brain. In Figure \ref{DTI_raw} the observed FA measurements at different tract locations in the CAA are shown, separately for cases and controls, i.e., individuals affected and non affected with MS. Each line in these plots represents an individual, and only the first visit is considered. Note that, in general, MS patients present lower FA measurements than healthy individuals. 

For illustration purposes, we present two different analyses and comparisons. We first focus our interest on the subgroup of individuals affected with MS. In this group, there are a total of $m = 99$ individuals, each with $s = 93$ FA measurements at different CAA tract locations. The SOP method is used to estimate the model described in Section \ref{OP_SSC} and presented in \cite{Djeundje10}. To compare results and computing times, the code associated with the paper by \cite{Djeundje10} is also tested. For this example both implementations take the advantage of the array structure of the data: Generalised Linear Array Models \citep[GLAM,][]{Currie2006} are used to efficiently compute the inner products for the Henderson equations (eqn. (\ref{MX:linearsystem})). Here, $43$ cubic B-spline basis are used for the population curve, and $23$ for the individual curves. This configuration gives rise to a model with $2320$ ($ = 43 + 23 \times 99$) coefficients (both random and fixed). SOP method needs about $150$ seconds to fit the model, and \cite{Djeundje10}'s code is $14$ times slower. We note that the computational time can be further improved if the sparse structure of the matrices involved in the model is exploited. Using the \texttt{R}-package \texttt{Matrix}, we are able to reduce the computing time using SOP to $35$ seconds. Figure \ref{DTI_results_pop} shows the estimated population effect using both approaches, that provide very similar results. $95\%$ pointwise confidence intervals are calculated by means of the full-sandwich standard errors proposed by \cite{Heckman13} but adapted to our case. Figure \ref{DTI_results_ind} shows, for several MS patients, the estimated (and observed) FA profiles. In terms of EDs, we obtain $35.03$ (out of $43$) for the population curve (including the unpenalised or fixed part), and a total of $2025.78$ (out of $2275$ ($= 23 \times 99 - 2$)) for all individual curves. We note that this total corresponds to the sum of the EDs associated with the variances $\sigma_2^2$ and $\sigma_3^2$ involved in the modelling of these individual curves (see expressions (\ref{SSC_P_ind}) and (\ref{SSC_G_ind}) for details). More precisely, using the SOP method we obtain an ED of $870.44$ for $\sigma_2^2$ and of $1155.34$ for $\sigma_3^2$. 

\begin{figure}
 \begin{center}
 \subfigure[SOP method]{\includegraphics[width=7cm]{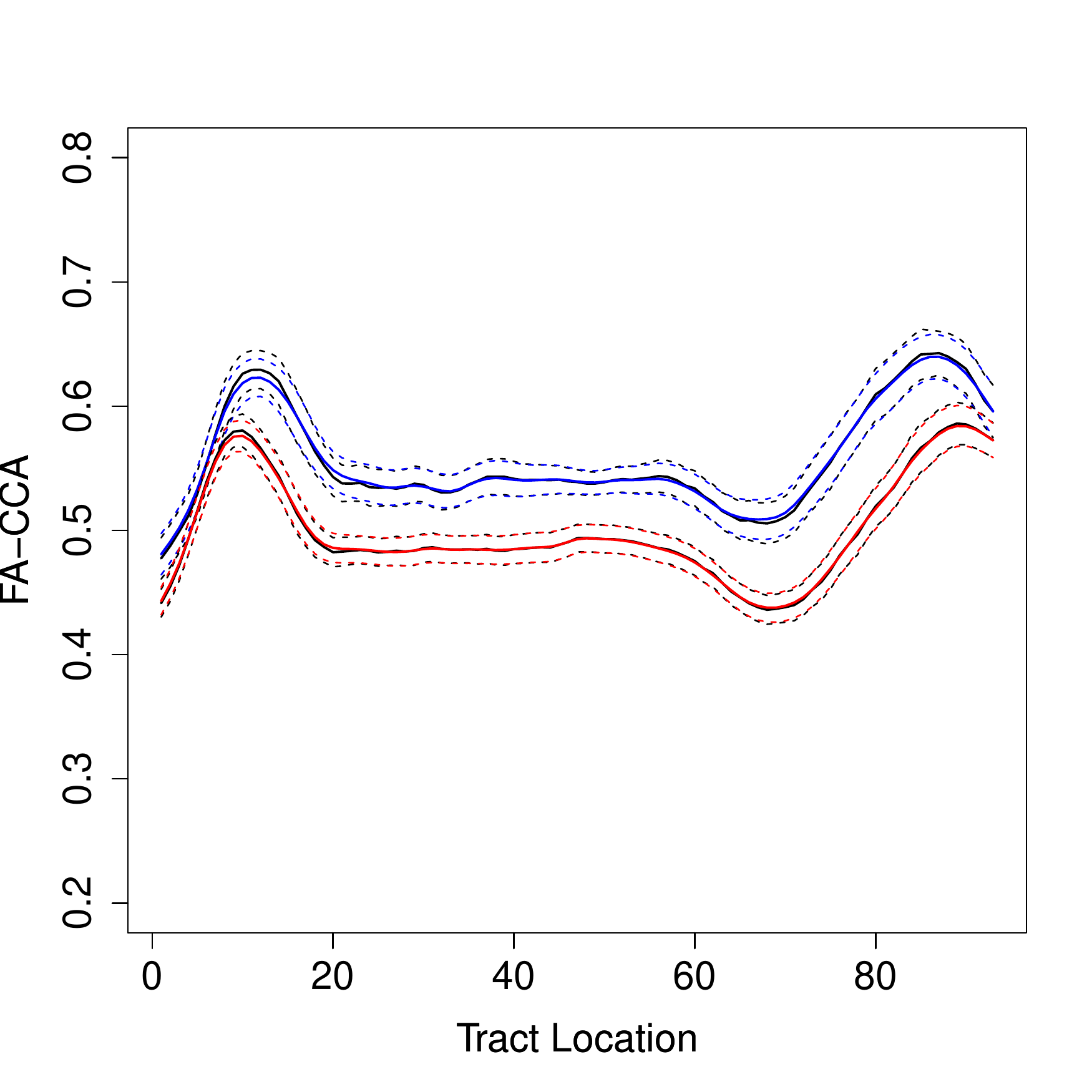}\label{DTI_results_pop_complete_SOP}}
 \subfigure[Functional approach]{\includegraphics[width=7cm]{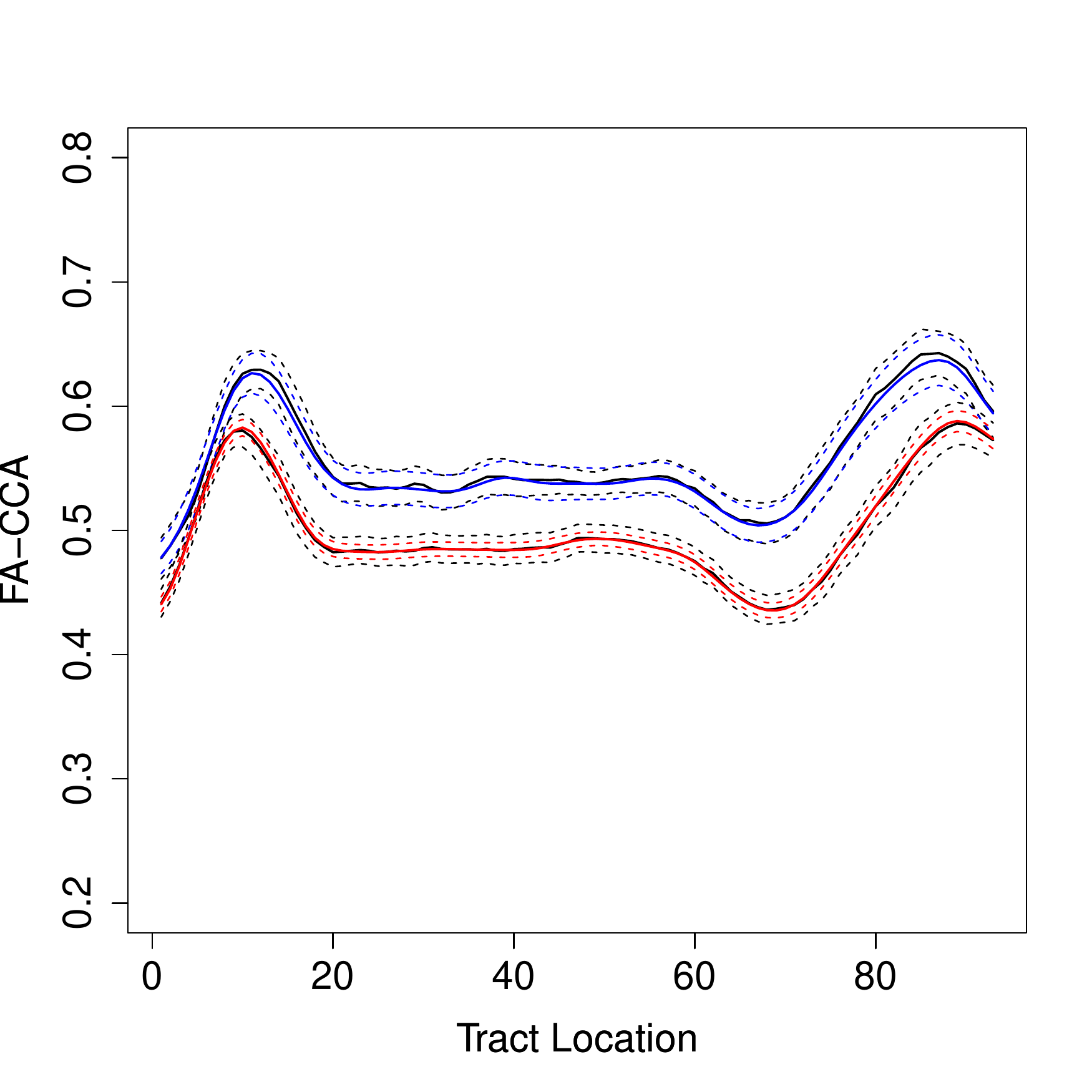}\label{DTI_results_pop_complete_functional}}
	\end{center}
		 \caption{For the diffusion tensor imaging scan data: estimated population (group) FA profiles. From left to right: results using the SOP method and the functional regression approach by \cite{Greven17}. Solid red line: MS patients. Solid blue line: controls. The dotted blue and red lines are the pointwise $95\%$ confidence intervals. For the SOP method confidence intervals are based on the full-sandwich standard errors proposed by \cite{Heckman13}. The solid back line is the observed mean and the dotted black lines are the empirical confidence intervals.}
\end{figure}

Our second analysis considers all individuals, cases and controls. The interest here is to compare the FA profiles at the first visit between these two groups. To that aim, the following factor-by-curve interaction model is considered. 
\[
y_{ij} = f_{z_j}\left(t_i\right) + g_j\left(t_i\right) + \varepsilon_{ij} \;\; 1\leq i \leq s,\;1\leq j \leq m,
\] 
where $z_j = 1$ if the $j$-th individual is affected by MS (case) and $z_j = 0$ otherwise (control). Note that this model assumes a different FA profile for each group. For this analysis, there a total of $m_0 = 42$ controls and $m_1 = 99$ MS patients ($m = m_0 + m_1 = 141$), and $s = 93$ different tract locations. A detailed description of the model can be found in Appendix \ref{AppC}. As for the first analysis $43$ cubic B-spline basis are used for the population curves (FA profiles), and $23$ for the individual curves, yielding a total of $3329$ ($ = 43 \times 2 + 23 \times 141$) coefficients. Using GLAM and sparse matrix techniques (\texttt{R}-package \texttt{Matrix}), the fit takes $65$ seconds. Figure \ref{DTI_results_pop_complete_SOP} shows the estimated FA profiles for both cases and controls, jointly with $95\%$ pointwise confidence intervals \citep{Heckman13}. The ED for the FA profile in controls is $32.21$ and in MS patients is $35.55$. In both cases we include the fixed part. Regarding the individual curves, we obtain a total ED of $2863.46$, the sum of $1263.26$ and $1600.20$. 

We compare the results with the functional regression model presented in the paper by \cite{Greven17} (see model (1.1) and Figure 2 in that paper). We would like to note that the P-spline model used in this paper was discussed in \cite{Durban17} as an competitive alternative to the functional approach by \cite{Greven17}. For the functional approach, we consider Gaussian homoskedastic errors, and, as suggested by the authors, $25$ cubic B-spline basis and a first order difference penalty, as well as $8$ functional principal components (FPC) functions. The code for fitting the model was kindly provided by the authors. Results are depicted in Figure \ref{DTI_results_pop_complete_functional}. As can be observed, both approaches provide very similar results. However, note that the pointwise $95\%$ confidence interval for the estimated FA profile in MS patients is narrower than the empirical confidence interval. This results can be explained by the (possibly wrong) assumption of Gaussian homoskedastic errors. In terms of computing times, the functional regression model needs $895$ seconds to be fitted (in contrast to $65$ seconds using SOP). We are aware that the computing times of both approaches (the SOP method and functional approach) are not fully comparable, since they assume different model specifications.

\section{Discussion}
This paper presents a new estimation method, called SOP, for (generalised) linear mixed models. The method is an extension of a previous proposal by \cite{Harville77}, and generalised by \cite{Schall1991} among others. In contrast to those previous approaches, the SOP method is suitable for models where the precision matrix is linear in the precision parameters. These precision matrices are common when penalised smooth models are reformulated as (generalised) linear mixed models. They appear when there are multiple quadratic penalties acting on the same regression coefficients. One special case are anisotropic tensor-product P-splines. This situation was discussed in the paper by \cite{MXRA2015} where the SAP algorithm was proposed. The present paper goes one step further and generalises SAP to a more general case. SOP is, as far as we know, the only method in this context where the variance parameters estimates involve ``partial'' effective degrees of freedom. As a consequence, the method provides, at convergence, the estimated effective partial degrees of freedom associated with each smooth/random component in the model. This is specially relevant when working with smoothing models, where the effective degrees of freedom of each smooth term gives important insights on the complexity of the fitted function. Furthermore, we show in the paper that the SOP method ensures non-negative estimates of the variance parameters, and discuss the conditions under which these estimates are strictly positive.

We present in the paper several situations in the context of penalised spline regression in which SOP represents a powerful alternative to existing estimation methods. In particular, we show the use of SOP in the case of spatially-adaptive P-splines and for the estimation of subject-specific curves in longitudinal data. We discuss several real data analyses dealing with these situations, and show the outperformance of SOP in terms of computing times. We use simple modelling situations with the aim of describing the method, models and examples in a detailed way, avoiding generalisations that could complicate the reading of the paper and obscure the simplicity of the proposal. However, there are several other fields of application of SOP. For instance, overlapping penalties appear in brain imaging research \citep{Karas2017} and derivative curve estimation \citep{simpkin2013}. In addition, the method can be used for more complex models including linear effects, (multidimensional) smooth functions, random Gaussian effects, etc. The use of other basis functions beyond B-splines is also possible as long as quadratic penalties are combined.

The proposal and examples presented in this paper pave the way for further research efforts. For instance, the approach discussed for adaptive P-splines is based on smoothing the locally varying smoothing parameter by means of B-splines. This implies that smoothness is solely controlled by the number of B-spline basis. The selection of the appropriate basis dimension may not be an easy task, with the undesirable consequence that if a large basis dimension is chosen (larger than needed), we may end up with a local linear fit. To reduce the impact of the basis dimension, we will explore the inclusion of a penalty on the coefficients (variance parameters) associated to the B-spline basis. This can be accomplished by means of a hierarchical structure for the random effects \cite[see, e.g.,][]{Krivobokova08}. Another challenging field is the study of suitable penalties and efficient estimation methods for adaptive P-splines in more than one dimension. Whereas some attempts have been done in two dimensions \cite[see, e.g.,][]{Crainiceanu07, Krivobokova08}, to the best of our knowledge the literature is lacking in three dimensional approaches (e.g., space and time). Some preliminary results using SOP are available at \cite{MXRA2016b}, but further work still needs to be done.

Finally, for variable selection there exists some works that propose sparse regression models using local quadratic approximations to the L1-norm adopting a penalised likelihood approach \citep[see][]{Fan2001,Hunter2005,Zou2008}. More recently, (generalised) linear mixed-effects approaches has been also proposed \cite[see][]{Taylor2012,Groll2014} allowing for the penalty to be estimated simultaneously with the variance parameters using REML. We intend to extend the SOP method in this direction. In conclusion, this paper opens up a pathway for a general estimating method allowing for both smoothing and variable selection in reasonable computing times.

The \texttt{R}-code used for the real data examples presented in Section \ref{examples} as well as an \texttt{R}-package implementing the SOP method for generic generalised linear mixed models, spatially-adaptive P-spline models and P-spline models for hierarchical curve data can be downloaded from \url{https://bitbucket.org/mxrodriguez/sop}.
\section*{Acknowledgements}
This research was supported by the Basque Government through the BERC 2018-2021 program and by Spanish Ministry of Economy and Competitiveness MINECO through BCAM Severo Ochoa excellence accreditation SEV-2013-0323 and through projects MTM2017-82379-R funded by (AEI/FEDER, UE) and acronym ``AFTERAM'', MTM2014-52184-P and MTM2014-55966-P. The MRI/DTI data were collected at Johns Hopkins University and the Kennedy-Krieger Institute. We are grateful to Pedro Caro and Iain Currie for useful discussions, to Martin Boer and Cajo ter Braak for the detailed reading of the paper and their many suggestions, and to Bas Engel for sharing with us his knowledge. This is a pre-print of an article published in \textit{Statistics and Computing}. The final authenticated version is available online at: \url{https://doi.org/10.1007/s11222-018-9818-2}.

\appendix
\section{Proof of Theorem \ref{the1}\label{AppA}}
\begin{proof}
We first note that the first-order partial derivatives of the (approximate) REML log-likelihood function can be expressed as \citep[see, e.g.,][]{MXRA2015}
\begin{equation*}
\frac{\partial{l}}{\partial{\sigma_{k_l}^2}} = - \frac{1}{2}trace\left(\boldsymbol{Z}^{\top}\boldsymbol{P}\boldsymbol{Z}\frac{\partial{\boldsymbol{G}}}{\partial{\sigma_{k_l}^2}}\right) + \frac{1}{2}\widehat{\boldsymbol{\alpha}}^{\top}\boldsymbol{G}^{-1}\frac{\partial{\boldsymbol{G}}}{\partial{\sigma_{k_l}^2}}\boldsymbol{G}^{-1}\widehat{\boldsymbol{\alpha}}.
\label{MX:rloglikder}
\end{equation*}
Given that $\boldsymbol{G}$ is a positive definite matrix, we have the identity
\[
\frac{\partial{\boldsymbol{G}}}{\partial{\sigma_{k_l}^2}} = - \boldsymbol{G}\frac{\partial{\boldsymbol{G}^{-1}}}{\partial{\sigma_{k_l}^2}}\boldsymbol{G},
\]
and thus
\[
\frac{\partial{\boldsymbol{G}}}{\partial{\sigma_{k_l}^2}} = \frac{1}{\sigma^{4}_{k_l}}\mbox{diag}\left(\boldsymbol{0}^{(1)},\boldsymbol{G}_k\boldsymbol{\Lambda}_{k_l}\boldsymbol{G}_k,\boldsymbol{0}^{(2)}\right), 
\]
where $\boldsymbol{0}^{(1)}$ and $\boldsymbol{0}^{(2)}$ are zero square matrices of appropriate dimensions. 

The first-order partial derivatives of the REML log-likelihood function are then expressed as
\begin{equation*}
2\frac{\partial{l}}{\partial{\sigma_{k_l}^2}} = - \frac{1}{\sigma_{k_l}^4}trace\left(\boldsymbol{Z}_k^{\top}\boldsymbol{P}\boldsymbol{Z}_k\boldsymbol{G}_k\boldsymbol{\Lambda}_{k_l}\boldsymbol{G}_k\right) + \frac{1}{\sigma_{k_l}^4}\widehat{\boldsymbol{\alpha}}_k^{\top}\boldsymbol{\Lambda}_{k_l}\widehat{\boldsymbol{\alpha}}_k.
\label{MX:rloglikder_2}
\end{equation*}
When the REML estimates are positive, they are obtained by equating the former expression to zero \cite[see, e.g.,][]{Engel1990}
\[
\frac{\widehat{\boldsymbol{\alpha}}^{\top}_k\boldsymbol{\Lambda}_{k_l}\widehat{\boldsymbol{\alpha}}_k}{trace\left(\boldsymbol{Z}_k^{\top}\boldsymbol{P}\boldsymbol{Z}_k\boldsymbol{G}_k\boldsymbol{\Lambda}_{k_l}\boldsymbol{G}_k\right)} = 1.
\]
We now multiply both sides with $\sigma_{k_l}^2$, and evaluate the left-hand side for the previous iterates and the right-hand side for the update, obtaining
\[
\widehat{\sigma}^{2}_{k_l} = \frac{\widehat{\boldsymbol{\alpha}}_k^{[t]\top}\boldsymbol{\Lambda}_{k_l}\widehat{\boldsymbol{\alpha}}_k^{[t]}}{trace\left(\boldsymbol{Z}_k^{\top}\boldsymbol{P}^{[t]}\boldsymbol{Z}_k\boldsymbol{G}_k^{[t]}\boldsymbol{\Lambda}_{k_l}\boldsymbol{G}_k^{[t]}\right)}\widehat{\sigma}_{k_l}^{2[t]} = \frac{\widehat{\boldsymbol{\alpha}}_k^{[t]\top}\boldsymbol{\Lambda}_{k_l}\widehat{\boldsymbol{\alpha}}_k^{[t]}}{trace\left(\boldsymbol{Z}_k^{\top}\boldsymbol{P}^{[t]}\boldsymbol{Z}_k\boldsymbol{G}_k^{[t]}\frac{\boldsymbol{\Lambda}_{k_l}}{\widehat{\sigma}_{k_l}^{2[t]}}\boldsymbol{G}_k^{[t]}\right)}.
\]
\end{proof}
\section{Proof of Theorem \ref{the2}\label{AppB}}
\begin{proof}
First let us recall some notation and introduce some needed results. We denote as $\boldsymbol{P} = \boldsymbol{V}^{-1} - \boldsymbol{V}^{-1}\boldsymbol{X}\left(\boldsymbol{X}^{\top}\boldsymbol{V}^{-1}\boldsymbol{X}\right)\boldsymbol{X}^{\top}\boldsymbol{V}^{-1}$, where $\boldsymbol{V} = \boldsymbol{R} + \boldsymbol{Z}\boldsymbol{G}\boldsymbol{Z}^{\top}$, $\boldsymbol{R} = \phi\boldsymbol{W}^{-1}$ with $\boldsymbol{W}$ being the diagonal weight matrix involved in the Fisher scoring algorithm.

Denote as $\mathcal{C}\left(\boldsymbol{A}\right)$ the linear space spanned by the columns of $\boldsymbol{A}$, and let $\boldsymbol{P}_{\boldsymbol{X}\boldsymbol{V}^{-1}} = \boldsymbol{X}\left(\boldsymbol{X}^{\top}\boldsymbol{V}^{-1}\boldsymbol{X}\right)^{-1}\boldsymbol{X}^{\top}\boldsymbol{V}^{-1}$ be the orthogonal projection matrix for $\mathcal{C}\left(\boldsymbol{X}\right)$ with respect to $\boldsymbol{V}^{-1}$. It is easy to show that
\[
\boldsymbol{P} = \boldsymbol{V}^{-1}\left(\boldsymbol{I}_n - \boldsymbol{P}_{\boldsymbol{X}\boldsymbol{V}^{-1}}\right) = \left(\boldsymbol{I}_n - \boldsymbol{P}_{\boldsymbol{X}\boldsymbol{V}^{-1}}\right)\boldsymbol{V}^{-1}\left(\boldsymbol{I}_n - \boldsymbol{P}_{\boldsymbol{X}\boldsymbol{V}^{-1}}\right).
\]
By Theorem 14.2.9 in \cite{Harville1997}, $\boldsymbol{P}$ is a (symmetric) positive semi-definite matrix. In addition,
\[
\boldsymbol{P}\boldsymbol{X} = \boldsymbol{0},
\]
and
\begin{align*}
\mbox{\mbox{rank}}(\boldsymbol{P}) & = \mbox{rank}\left(\boldsymbol{V}^{-1}\left(\boldsymbol{I}_n - \boldsymbol{P}_{\boldsymbol{X}\boldsymbol{V}^{-1}}\right)\right)\\
& = \mbox{rank}\left(\left(\boldsymbol{I}_n - \boldsymbol{P}_{\boldsymbol{X}\boldsymbol{V}^{-1}}\right)\right)\\
& = n - \mbox{rank}\left(\boldsymbol{P}_{\boldsymbol{X}\boldsymbol{V}^{-1}}\right)\\
& = n - \mbox{rank}\left(\boldsymbol{X}\right).
\end{align*}
Thus,
\begin{equation}
ker\left(\boldsymbol{P}\right) = \mathcal{C}\left(\boldsymbol{X}\right),
\label{k_p_c_x}
\end{equation}
i.e., $\boldsymbol{P}\boldsymbol{x} = \boldsymbol{0}$ if and only if $\boldsymbol{x}$ is in $\mathcal{C}\left(\boldsymbol{X}\right)$.

Let $\boldsymbol{\Lambda}_{k_l} = \boldsymbol{U}\boldsymbol{\Sigma}\boldsymbol{U}^{\top}$ be the eigen value decomposition of $\boldsymbol{\Lambda}_{k_l}$. Note that $\boldsymbol{\Lambda}_{k_l} = \boldsymbol{U}_{+}\boldsymbol{\Sigma}_{+}\boldsymbol{U}_{+}^{\top}$, where $\boldsymbol{U}_{+}$ and $\boldsymbol{\Sigma}_{+}$ are the sub-matrices corresponding to the non-zero eigenvalues. Then
\[
\widehat{\boldsymbol{\alpha}}_k^{\top}\boldsymbol{\Lambda}_{k_l}\widehat{\boldsymbol{\alpha}}_k = \widehat{\boldsymbol{\alpha}}_k^{\top}\boldsymbol{U}_{+}\boldsymbol{\Sigma}_{+}\boldsymbol{U}_{+}^{\top}\widehat{\boldsymbol{\alpha}}_k = \boldsymbol{z}^{\top}\boldsymbol{P}\boldsymbol{Z}_k\boldsymbol{G}_k\boldsymbol{U}_{+}\boldsymbol{\Sigma}_{+}\boldsymbol{U}_{+}^{\top}\boldsymbol{G}_k\boldsymbol{Z}_k^{\top}\boldsymbol{P}\boldsymbol{z} \geqslant 0,
\]
with equality holding if and only if $\boldsymbol{U}_{+}^{\top}\boldsymbol{G}_k\boldsymbol{Z}_k^{\top}\boldsymbol{P}\boldsymbol{z} = \boldsymbol{0}$ (since $\boldsymbol{\Sigma}_{+}$ is positive definite). Thus, using result (\ref{k_p_c_x}), the equality holds if $\boldsymbol{z}$ is in $\mathcal{C}\left(\boldsymbol{X}\right)$ or $\mathcal{C}\left(\boldsymbol{Z}_k\boldsymbol{G}_k\boldsymbol{U}_{+}\right) \subset \mathcal{C}\left(\boldsymbol{X}\right)$. By Lemma 4.2.2 and Corollary 4.5.2 in \cite{Harville1997}, we have
\[
\mathcal{C}\left(\boldsymbol{Z}_k\boldsymbol{G}_k\boldsymbol{U}_{+}\right) = \mathcal{C}\left(\boldsymbol{Z}_k\boldsymbol{G}_k\boldsymbol{\Lambda}_{k_l}\right) \subset \mathcal{C}\left(\boldsymbol{X}\right) \iff \mbox{rank}\left(\boldsymbol{Z}_k\boldsymbol{G}_k\boldsymbol{\Lambda}_{k_l},\boldsymbol{X}\right) = \mbox{rank}\left(\boldsymbol{X}\right).
\]

Regarding the denominator of the REML-based estimates updates, we follow a similar reasoning. Using Corollary 14.7.5 (and Theorem 14.2.9) in \cite{Harville1997}, we have
\begin{align*}
trace\left(\boldsymbol{Z}_k^{\top}\boldsymbol{P}\boldsymbol{Z}_k\boldsymbol{G}_k\boldsymbol{\Lambda}_{k_l}\boldsymbol{G}_k\right) = trace\left(\boldsymbol{U}_{+}^{\top}\boldsymbol{G}_k\boldsymbol{Z}_k^{\top}\boldsymbol{P}\boldsymbol{Z}_k\boldsymbol{G}_k\boldsymbol{U}_{+}\boldsymbol{\Sigma}_{+}\right) \geqslant 0,
\end{align*}
with equality holding if and only if $\boldsymbol{U}_{+}^{\top}\boldsymbol{G}_k\boldsymbol{Z}_k^{\top}\boldsymbol{P}\boldsymbol{Z}_k\boldsymbol{G}_k\boldsymbol{U}_{+} = \boldsymbol{0}$. Again, this equality holds if and only if $\mathcal{C}\left(\boldsymbol{Z}_k\boldsymbol{G}_k\boldsymbol{U}_{+}\right) \subset \mathcal{C}\left(\boldsymbol{X}\right)$ (i.e., $\iff \mbox{rank}\left(\boldsymbol{Z}_k\boldsymbol{G}_k\boldsymbol{\Lambda}_{k_l},\boldsymbol{X}\right) = \mbox{rank}\left(\boldsymbol{X}\right)$).
\end{proof}
\section{Estimating algorithm}\label{algorithm}
This Appendix summarises the steps of the estimating algorithm for model (\ref{mm_equation}) based on the SOP method:
\begin{description}
\item[Initialise.] Set initial values for $\widehat{\boldsymbol{\mu}}^{[0]}$ and the variance parameters $\widehat{\sigma}_{k_l}^{2[0]}$ ($l = 1,\ldots,p_k$ and $k = 1, \ldots, c$). In those situations where $\phi$ is unknown, establish an initial value for this parameter, $\widehat{\phi}^{[0]}$. Set $t = 0$. 
\item[Step 1.] Construct the \textit{working} response vector $\boldsymbol{z}$ and the matrix of weights $\boldsymbol{W}$ as follows
\[
z_i = g(\widehat{\mu}_i^{[t]}) + (y_i - \widehat{\mu}_i^{[t]})g^{\prime}(\widehat{\mu}_i^{[t]}),
\]
\[
w_{ii} = \left\{g'(\widehat{\mu}_i^{[t]})^2\nu(\widehat{\mu}_i^{[t]})\right\}^{-1}.
\]
\begin{description}
\item[Step 1.1.] Given the initial \textit{estimates} of variance parameters, estimate $\boldsymbol{\alpha}$ and $\boldsymbol{\beta}$ by solving
\begin{equation}
\underbrace{
\begin{bmatrix}
\boldsymbol{X}^{\top}{\boldsymbol{R}^{[t]}}^{-1}\boldsymbol{X} & \boldsymbol{X}^{\top}{\boldsymbol{R}^{[t]}}^{-1}\boldsymbol{Z} \\
\boldsymbol{Z}^{\top}{\boldsymbol{R}^{[t]}}^{-1}\boldsymbol{X} & \boldsymbol{Z}^{\top}{\boldsymbol{R}^{[t]}}^{-1}\boldsymbol{Z} + {\boldsymbol{G}^{[t]}}^{-1}
\end{bmatrix}}_{\boldsymbol{C}^{[t]}}
\begin{bmatrix} 
\boldsymbol{\widehat{\beta}}\\
\boldsymbol{\widehat{\alpha}}
\end{bmatrix}
=
\begin{bmatrix}
\boldsymbol{X}^{\top}{\boldsymbol{R}^{[t]}}^{-1}\boldsymbol{z}\\
\boldsymbol{Z}^{\top}{\boldsymbol{R}^{[t]}}^{-1}\boldsymbol{z}
\end{bmatrix},
\label{MX:linearsystem}
\end{equation}
where $\boldsymbol{R}^{[t]} = \widehat{\phi}^{[t]}\boldsymbol{W}^{-1}$. Let $\boldsymbol{\widehat{\alpha}}^{[t]}$ and $\boldsymbol{\widehat{\beta}}^{[t]}$ be these estimates.
\item[Step 1.2.] Update the variance parameters as follows
\begin{equation*}
\widehat{\sigma}^{2}_{k_l} = \frac{\widehat{\boldsymbol{\alpha}}_k^{{[t]}\top}\boldsymbol{\Lambda}_{k_l}\widehat{\boldsymbol{\alpha}}_k^{[t]}}{\mbox{ED}_{k_l}^{[t]}},
\end{equation*}
and, when necessary,
\begin{equation*}
\widehat{\phi} = \frac{\left(\boldsymbol{z}-\boldsymbol{X}\boldsymbol{\widehat{\beta}}^{[t]}-\boldsymbol{Z}\boldsymbol{\widehat{\alpha}}^{[t]}\right)^{\top}\boldsymbol{W}\left(\boldsymbol{z}-\boldsymbol{X}\boldsymbol{\widehat{\beta}}^{[t]}-\boldsymbol{Z}\boldsymbol{\widehat{\alpha}}^{[t]}\right)}{n - \mbox{rank}(\boldsymbol{X}) - \sum_{k=1}^c\sum_{l=1}^{p_k}\mbox{ED}_{k_l}^{[t]}},
\end{equation*}
with
\[
\mbox{ED}_{k_l}^{[t]} = trace\left(\left(\boldsymbol{G}^{[t]}_k - {\boldsymbol{C}^{[t]}}^{*}_{kk}\right)\frac{\boldsymbol{\Lambda}_{k_l}}{{\widehat{\sigma}_{k_l}^{2[t]}}}\right).
\]
Recall that ${\boldsymbol{C}^{[t]}}^{*}$ denotes the inverse of $\boldsymbol{C}^{[t]}$ in (\ref{MX:linearsystem}), and ${\boldsymbol{C}^{[t]}}^{*}_{kk}$ is that partition of ${\boldsymbol{C}^{[t]}}^{*}$ corresponding to the $k$-th random component $\boldsymbol{\alpha}_k$.
\item[Step 1.3.] Repeat Step 1.1 and Step 1.2, replacing $\widehat{\sigma}_{k_l}^{2[t]}$, and, if updated, $\widehat{\phi}^{[t]}$, by $\widehat{\sigma}_{k_l}^{2}$ and $\widehat{\phi}$, until convergence. For the examples presented in Section \ref{examples}, we use the REML deviance as the convergence criterion. 
\end{description}
\item[Step 2.] Repeat Step 1, replacing the variance parameters and the model's fixed and random effects (and thus $\widehat{\boldsymbol{\mu}}^{[t]}$) by those obtained in the last iteration of Steps 1.1 - Step 1.3, until convergence.
\end{description}
\section{Factor-by-curve hierarchical curve model\label{AppC}}
This appendix describes in detail the factor-by-curve interaction model discussed in Section \ref{DTI_example}, i.e., 
\[
y_{ij} = f_{z_j}\left(t_i\right) + g_j\left(t_i\right) + \varepsilon_{ij} \;\; 1\leq i \leq s,\;1\leq j \leq m,
\] 
where $z_j = 1$ if the $j$-th individual is affected by MS (case) and $z_j = 0$ otherwise (control). Let's order the data with the observations on controls first, followed by observations on MS patients. In matrix notation, the model can be expressed as
\begin{equation}
\boldsymbol{y} = [\boldsymbol{Q}\otimes\boldsymbol{B}]\boldsymbol{\theta} + [\boldsymbol{I}_{m}\otimes\breve{\boldsymbol{B}}]\breve{\boldsymbol{\theta}} + \boldsymbol{\varepsilon},
\label{SSC_inter_model_pop}
\end{equation}
with $\boldsymbol{B}$, $\breve{\boldsymbol{B}}$, $\breve{\boldsymbol{\theta}}$ and $\boldsymbol{\varepsilon}$ as defined in Section \ref{OP_SSC}. Matrix $\boldsymbol{Q}$ is any suitable contrast matrix of dimension $m \times 2$, where $m = m_0 + m_1$, with $m_0$ being the number of controls and $m_1$ the number of MS patients. For our application, we consider
\[
\boldsymbol{Q} = \begin{pmatrix}
\boldsymbol{1}_{m_0} & \boldsymbol{0}_{m_0}\\
\boldsymbol{0}_{m_1} & \boldsymbol{1}_{m_1}
\end{pmatrix},
\]
and a different amount of smoothing is assumed for $f_{0}$ and $f_{1}$, i.e., the penalty matrix acting over the vector of coefficients $\boldsymbol{\theta}$ is of the form
\[
\boldsymbol{P} = \begin{pmatrix}
\lambda_{1}\boldsymbol{D}_q^{\top}\boldsymbol{D}_q & \boldsymbol{0}_{c \times c}\\
\boldsymbol{0}_{c \times c} & \lambda_{2}\boldsymbol{D}_q^{\top}\boldsymbol{D}_q
\end{pmatrix}. 
\]
The reformulation as a mixed model can be done in a similar fashion to that described in Section \ref{OP_SSC}, with, in this case
\begin{align*}
\boldsymbol{X} = & [\boldsymbol{Q}\otimes\boldsymbol{B}\boldsymbol{U}_{0}],\\
\boldsymbol{Z} = & [\boldsymbol{Q}\otimes\boldsymbol{B}\boldsymbol{U}_{+}:\boldsymbol{I}_m\otimes\breve{\boldsymbol{B}}],\\
\end{align*}
and
\[
\boldsymbol{G}^{-1} = 
\begin{pmatrix}
\sigma_{1}^{-2}\boldsymbol{\Sigma}_{+} & \boldsymbol{0}_{d \times d} & \boldsymbol{0}_{d\times\left(m \breve{d}\right)}\\
\boldsymbol{0}_{d\times d} & \sigma_{2}^{-2}\boldsymbol{\Sigma}_{+} & \boldsymbol{0}_{d\times\left(m \breve{d}\right)}\\
\boldsymbol{0}_{\left(m \breve{d}\right)\times d}& \boldsymbol{0}_{\left(m \breve{d}\right)\times c} & \boldsymbol{I}_{m}\otimes\breve{\boldsymbol{G}}^{-1}
\end{pmatrix},
\]
where
\[
\breve{\boldsymbol{G}}^{-1} = \sigma_3^{-2}\breve{\boldsymbol{D}}_{\breve{q}}^{\top}\breve{\boldsymbol{D}}_{\breve{q}} + \sigma_4^{-2}\boldsymbol{I}_{\breve{d}}. 
\]
\bibliographystyle{hapalike}
\bibliography{kk}
\end{document}